\documentclass[%
reprint,
amsmath,amssymb,
aps,
]{revtex4-1}

\usepackage{graphicx}
\usepackage{dcolumn}
\usepackage{bm}
\usepackage{hyperref}
\newif\ifhyper
\hypertrue
\ifhyper
\hypersetup{
	citecolor = {red},
	colorlinks = {true}, 
	linkcolor = {blue},
	urlcolor = {blue} 
}
\fi

\begin{document}
\newcommand\Aplocal{C}
\newcommand\Apmiccom{E}
\newcommand\Aplocalopt{C.2}
\newcommand\Aplocalheto{C.3}
\newcommand\Aplocalhet{C.4}
\newcommand\Apsim{F}
\newcommand\Apsimq{F.1}
\newcommand\Apsimnc{F.2}
\newcommand\Apglob{D}

\title{Evolutionary value of collective sensing}

\author{Mohammad Salahshour}
\email{salahshour\_mohammad@physics.sharif.edu}
\author{Shahin Rouhani}%
 \email{srouhani@sharif.ir}
\affiliation{%
  Department of Physics, Sharif University of Technology, P.O. Box 11165-9161, Tehran, Iran\\}%




\date{\today}

\begin{abstract}
We propose a mathematical model for collective sensing in a population growing in a stochastically varying environment. In the population, individuals use an information channel for sensing the environment, and two channels for signal production and comprehension to communicate among themselves. We show that existence of such system has a positive effect on population growth, hence can have a positive evolutionary effect. We show that the gain in growth due to the collective sensing is related to information theoretic entities, which can be considered as the information content of this system from the environment. We further show that heterogeneity in communication resulted from network or spatial structure increases growth. We compute the growth rate of a population residing on a lattice and show that growth rate near the maximum noise level in observation or communication, increases exponentially as noise decreases. This exponential effect makes the emergence of collective observation an easy outcome in an evolutionary process. Furthermore, we are able to quantify interesting effects such as accelerated growth, and simplification of decision making due to information amplification by communication. Finally, we show that an amount of noise in representation formation has more disadvantageous effect compared to the same noise in signal production.
\begin{description}
	\item[Keywords] Collective Sensing, Communication, Information, Growth Rate.
\end{description}
\end{abstract}

\pacs{Valid PACS appear here}
\keywords{Collective Sensing | Communication | Information | Growth Rate |}

\maketitle

\section{Introduction}
Acquisition of information from environmental conditions is vital for populations living in stochastically varying environment, and evolution has developed many mechanisms to accomplish this task \cite{Dall,tkavcik,Kussell,rvoire,Mayer,Skanata,Kobayashi}. One of the most astonishing ones is collective sensing, in which individuals belonging to a community make observations of the environment and communicate their observations 
by production, transmission and comprehension of signals. Individuals use the information they have reached, either through direct observation or through communication, to decide upon their strategy. Examples of such phenomena are many. Bacteria communicate using chemical signals, in what is known as quorum sensing, to infer their changing environment and use this information for informed decision making \cite{Waters,Cornforth,Popat}. It has been noted that quorum sensing bacteria can optimize growth and invade other non communicating phenotypes \cite{Bruger}. Individual cells in a multicellular organism, use a rich set of chemical signals to divide labour, coordinate behaviour and respond to the changes of the outside world \cite{Alon}. Insects in social and eusocial species, as well as many group living large animals, use many sorts of signals to communicate and exchange information, and use this information for decision making in responding to the environmental challenges \cite{Leonhardt,Tereshko,Brumm}.
Finally, Human language, seen as a system of signs which individuals use to communicate representations, is judged to have similarities with these primitive communication systems \cite{Martin,Komarova,Hendriks}\\
Given the prevalence of such collective sensing systems in biological populations, an important question is what is the evolutionary advantage of collective sensing and how it emerges in an evolutionary process? To answer this question, here we consider a model in which the growth rate of a population living in a stochastically varying environment, is affected by its capability to select the best response to environmental conditions. The population uses a collective sensing system (CSS) to reach the information needed in choosing a response. We quantify the effect of this system on the fitness of the population by showing that when operating optimally, the effect of a CSS on growth rate can be quantified by information theoretic terms which can be construed as its information content from environment. Depending on the structure of communications and the properties of the CSS, this term is either a generalization of, approaches, or is equal to the mutual information between the environmental process and the set of representations of the CSS.\\
As such CSSs generally operate under a noisy condition, it is important to assess their performance as a function of noise level present in their faculties, specifically, in representation formation, and signal production and comprehension. We will tackle this issue by showing that the growth rate increases exponentially, as the level of noise in the CSS faculties is decreased near the maximum noise level. The exponential increase in the performance of the CSS as the level of noise is reduced, makes the emergence of such CSSs an easy outcome in an evolutionary process, as starting from a maximally noisy, random observation and communication system, small noise reduction steps can have significant fitness effects. This can explain the prevalence of such CSSs in the biological world. Furthermore, we show that communication amplifies the positive effect of noise reduction in observation. This implies, there should be higher pressure for noise reduction in sensory observations in populations who communicate compared to those who do not.\\
We will also characterize the effect of communication network on growth, by showing that heterogeneity in communication resulted from network or spatial structure, increases growth by producing diversity. We compare the effect of noise in representation of the environment and production of signals, and show that the detrimental effect of some amount of noise in representation is higher compared to the case when the same amount of noise is present in production. Finally, we will be able to quantify interesting effects such as accelerated growth or simplification of decision making resulted from amplification of information through communication.\\
\section{The Model}
\subsection{Population growth Model}
We consider a simplified model of population growth in which a population evolves in a stationary stochastically varying environment $\epsilon_t$. The environment can assume $n$ states $\epsilon$, with the stationary marginal probability distribution $p_{\epsilon}$. An individual living in such an environment has $n$ different strategies, each a proper response to an environmental state. We denote the proper strategy in environment state $\epsilon$ by $s_{\epsilon}$.
We show the total population size at time $t$ by $N_t$, and the sub-population adopting strategy $s$ at time $t$ by $N_{t,s}$. The growth of the population in general should be proportional to the amount of resources correctly devoted to deal with the current environmental challenges. As a sub population of $N_{t,s_{\epsilon_t}}$ of the total population devote their effort correctly to deal with the current environment, it is reasonable to set the growth rate proportional to $N_{t,s_{\epsilon_t}}$. We take the proportionality constant to depend on the environment, and denote it by $w_{\epsilon_t}$. In addition, for simplicity we assume discrete growth. Given all these, the population evolution equation becomes:
\begin{align}
N_{t+1}=w_{\epsilon_t}N_{t,s_{\epsilon_t}}.
\label{popev}
\end{align}
We define the long term growth rate or Lyapunov exponent as \cite{Kussell,Rivoire,Donaldson}:
\begin{align}
\Lambda=\lim_{T \to \infty}\frac{1}{T}\log(N_T/N_0).
\label{Lyap}
\end{align}
\subsection{CSS Model}
To give a mathematical model for a CSS, inspired by biological examples, we consider a language-like system in which individuals acquire information from the world in form of internal representations or senses, and exchange their information using signals. Individuals use the information they have obtained 
to decide upon their strategy to respond to environment.
Therefore, we define a CSS as composed of $n_r$ representations, each intended to represent a particular environmental or world state, $n_{\sigma}$ signals which are used to signify representations, and $n_s$ actions or strategies. In addition a CSS is identified by three probability transition matrices: a representation matrix $R(r|\epsilon)$ which is the probability of using representation $r$ for world state $\epsilon$, a production matrix $G(\sigma|r)$, which is the probability of using signal $\sigma$ for representation $r$, a comprehension matrix $C(r|\sigma)$ which is the probability of interpreting representation $r$ when receiving signal $\sigma$.
\\As a result of observation and communication, individuals form expectations of the state of the world, which is modelled here as a set of representations $\pmb{r}$ that an individual reaches and stores as an internal state. Here, a bold letter $\pmb{r}$ means a set of representations $r$. An individual acts upon the environment based on her internal state. This is accomplished by a decision or action matrix $A(s|\pmb{r})$, which is the probability of choosing strategy $s$ given the internal state $\pmb{r}$.\\
The dynamics of the model can be thought of as follows. In each environmental state $\epsilon$, some or all of the individuals observe the world state using the channel $R(r|\epsilon)$. The most natural model is the one in which, in each environmental state, each individual makes an observation with probability $q$. As a result of observation, each observer obtains a representation $r$ drawn according to $R(r|\epsilon)$, given the environmental state $\epsilon$. Then, those who have made an observation, transmit a signal to a set of other individuals. We will consider models where a transmitter transmits her signal globally to a non-negligible fraction of the population, or when she transmits her signal to her local neighbourhood. To transmit a signal, a transmitter produces a signal $\sigma$, given her representation $r$ according to the production matrix $G(\sigma|r)$. Finally, the receiver of a signal $\sigma$, interprets the signal as referring to representation $r^{\rho}$ according to her comprehension matrix $C(r^{\rho}|\sigma)$. Here, the upper index $\rho$ stands for receiver. As a result, each individual reaches an internal state $\pmb{r}^{\rho}$, which is composed of all the internal representations that she obtains through observation or communication, and chooses strategy s with probability $A(s|\pmb{r}^{\rho})$.\\
\begin{figure}
	\includegraphics[width=1\linewidth]{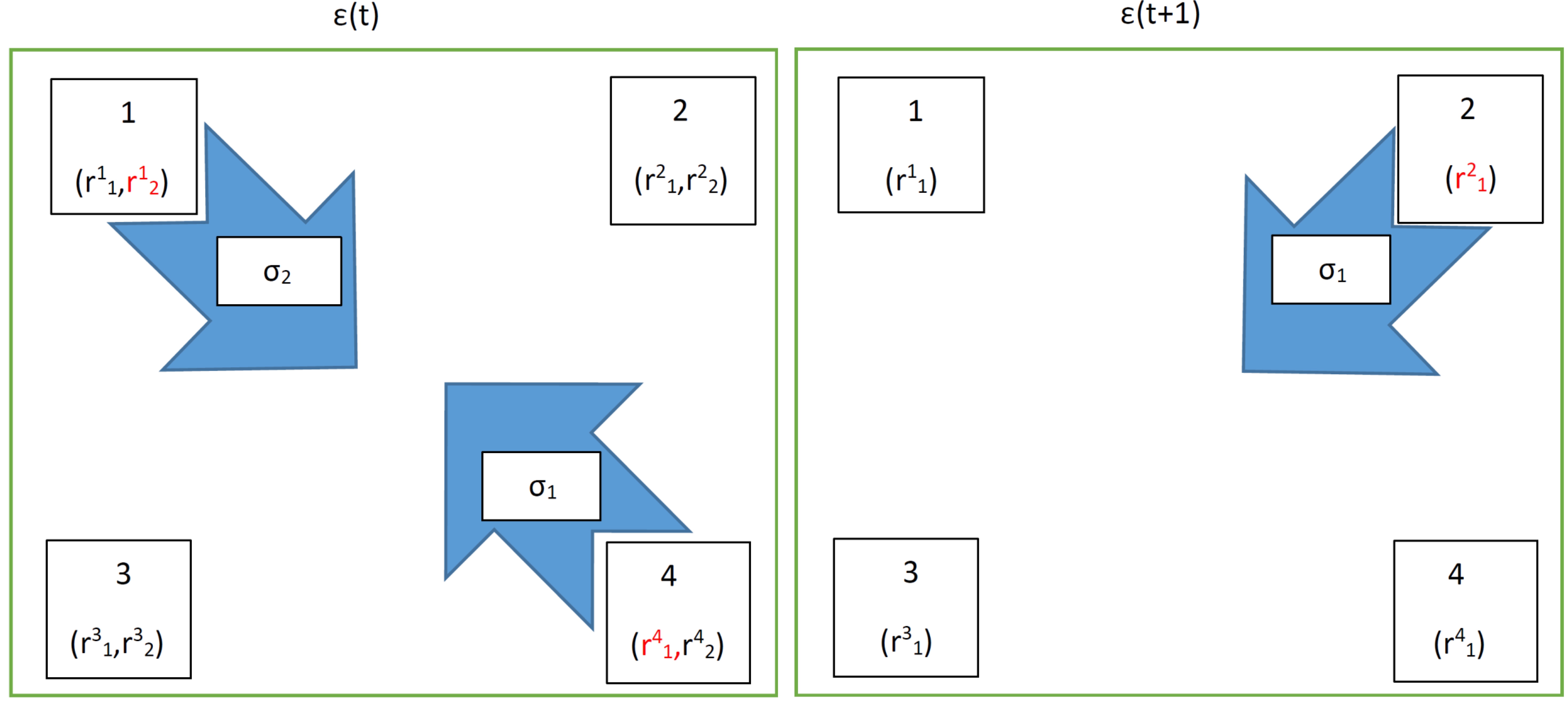}
	\caption{Dynamics of the model. In each environment $\epsilon_t$ some agents detect the world state and send a signal to a set of receivers, which can be the whole population (global communication), or a local neighbourhood (local communication). Here in time $t$, $\epsilon_t$ occurs and agents $1$ and $4$ make an observation using $R(r|\epsilon_t)$, and reach respectively the representations $r^1_2$ and $r^4_1$. The upper index shows the agent number and lower index is the representation number. Those who have made the observation send a signal using $G(\sigma|r)$ to others. Here agents $1$ and $4$ send respectively signals $\sigma_1$ and $\sigma_2$. Those who receive the signal interpret it as referring to a representation using $C(r|\sigma)$. Here the signal is transmitted to the whole population and consequently the whole population reach an internal state which is composed of $2$ representations. In time $t+1$ only one agent (agent $2$) has made an observation and transmit a signal to the rest of the population, consequently the internal state of the population is composed of $1$ representation.}
	\label{figPLM}
\end{figure}
Assuming a CSS is given by the matrices $(R,G,C)$, we show the growth rate of a population having this CSS by $\Lambda_{(R,G,C)}$, and its optimal growth rate by $\Lambda_{(R,G,C)}^*$. In order to have a measure of the gain in fitness or long term growth rate of a population due to having a CSS, we ask, assuming a population is equipped with a CSS, what would be its gain compared to the same population without a CSS? We call this gain, the fitness value of the CSS, $V_{CSS}$ \cite{rvoire, Rivoire, Donaldson}. Thus the value of a CSS is $V_{(R,G,C)}=\Lambda_{(R,G,C)}^*-\Lambda_{0}^*$. Where $\Lambda_0^*$ is the optimal growth of the same population without a CSS. To compute $\Lambda_0^*$, we consider a population who lacks a CSS. In this case individuals have to make decisions in absence of any information provided by the CSS and their action matrix will be $A(s|.)$. Here a dot means an empty set. Through this paper we make the assumption that the population size is large. With this assumption we have $N_{t,s_{\epsilon_t}}=A(s_{\epsilon_t}|.)N_t$. Putting this in the population evolution equation Eq. (\ref{popev}), Iterating it from time $0$ to $T-1$ and dividing by initial population size, taking the logarithm, dividing by $T$, taking the large time limit and using the stationarity of the process to replace time summation with a summation over states, we derive for the long term growth rate as defined in Eq. (\ref{Lyap}):
\begin{align}
\Lambda_{0}=\sum_{\epsilon}p_\epsilon log(w_{\epsilon}A(s_{\epsilon}|.)).
\label{basegrowth}
\end{align}
By optimizing this subject to the constraint $\sum_{\epsilon}A(s_{\epsilon}|.)=1$, We have for the optimal action matrix $A(s_{\epsilon}|.)=p_{\epsilon}$, and for the optimal growth rate:
\begin{align}
\Lambda_{0}^*=
\sum_{\epsilon=1}^{n}p_{\epsilon}log(w_{\epsilon})-H_{p_{\epsilon}}(\epsilon).
\label{lambda1}
\end{align}
In which $H_{p_{\epsilon}}(\epsilon)$ is the entropy of the environmental stochastic process and quantifies the amount of uncertainty in, or the information content of the environmental process \cite{Cover}. We see that this uncertainty reduces the growth of a population who does not have knowledge of the process.\\
Here are some notes on notation. We show sets of elements $x$ by bold lower case letters $\pmb{x}$. We use bold upper case letters to denote concatenation of matrices as $\pmb{X}=\prod X$. We normally drop convolution signs. Writing matrices next to each other implies convolution. In general signals have a propagation length scale beyond which can not propagate. When the propagation length scale is finite, a signal reaches a local neighbourhood around the transmitter. In infinite population size, which is the limit we work in here, this is a negligible fraction of the population. We call this regime local communication. In contrast, in global communication, a signal reaches whole or a finite fraction of the population. This can result from infinite propagation length scale. The solution of the model differs in the two cases and we begin with the former.
\section{Local communication}
\label{localmain}
\subsection{The value of CSS with local communication}
We consider here local communication in which a transmitter transmits her signal locally to a small set of receivers. Such localities can result from a characteristic propagation length scale of signals in a spatial structure. Another possibility is that local communication results from a network structure on which the population resides, such that individuals communicate only with their neighbours on the network. The calculations in this section are detailed in Appendix \Aplocal.\\
To work with a specific model, we assume, in each environmental state, each individual makes an observation with probability $q$ using $R(r|\epsilon)$. Reaching internal representation $r$, she transmits a signal to her local neighbourhood to inform them of the result of her observation. The signal is produced according to $G(\sigma|r)$. As a result of local communications, individuals receive different profile of signals $\pmb{\sigma}_l$. A lower index $l$ here, refers to local communication. We can define a probability distribution $\pmb{N}_{l}(\pmb{\sigma}_l|\epsilon)$ which gives the probability that an individual receives a profile $\pmb{\sigma}_l$ in the environmental state $\epsilon$. The form of this probability distribution can be worked out given the assumptions about the structure of interactions. The results of this section are not dependent on the detailed structure of interactions and the exact form of this distribution. However, two simple such models are introduced in Appendix \Apmiccom.\\
With these assumptions, in large population size limit, a fraction $q$ of the population make an observation and in addition, obtain a profile of internal representations $\pmb{r}^{\rho}$, through communication according to $\pmb{C}_l\pmb{N}_l(\pmb{r}_l^{\rho}|\epsilon_t)$. Thus, their internal states become of the form $(r,\pmb{r}^{\rho})$. Finally, they choose their strategy according to $A_1(s|r,\pmb{r}^{\rho})$. The rest, do not observe the world directly and only obtain a profile of internal representation $\pmb{r}^{\rho}$ according to $\pmb{C}_l\pmb{N}_l(\pmb{r}_l^{\rho}|\epsilon_t)$. Consequently, their internal state becomes of the form $(\pmb{r}^{\rho})$. They choose their strategy according to $A_2(s|\pmb{r}^{\rho})$. Given all these, it is easy to see that the population evolution equation becomes $N_{t+1}= \sum_{\pmb{r}_l^{\rho}} w_{\epsilon_{t}}(\sum_{  r}qA_1(s_{\epsilon_t}|r,\pmb{r}_l^{\rho})R(r|\epsilon_t)+(1-q)A_2(s_{\epsilon_t}|\pmb{r}^{\rho}_l))\pmb{C}_l\pmb{N}_l(\pmb{r}^{\rho}_l|\epsilon_t)N_t$. As before, iterating this equation from time $0$ to $T-1$, dividing by initial population size, taking the logarithm and dividing by $T$, and using the stationarity of the process to replace time summation with a summation over states, we derive for the growth rate:
\begin{align}
&\Lambda_{(A,G,C)}^{l}=\sum_{\epsilon}p_{\epsilon}\log(w_{\epsilon}A\pmb{N}_q\pmb{N}_l(s_{\epsilon}|\epsilon)).
\label{het6main}
\end{align}
Where we have defined $\pmb{N}_q(z,\pmb{r}^{\rho}|\epsilon,\pmb{\sigma})=(qR(z|\epsilon)+(1-q)\delta_{z,.})\pmb{C}(\pmb{r}^{\rho}|\pmb{\sigma}_l)$. Here, $z$ is a random variable which can be either $r$ or an empty set $.$, and $R(z=r|\epsilon)=R(r|\epsilon)$ and $R(z=.|\epsilon)=0$. Here, we have also introduced a delta function $\delta_{z,.}$ which is $1$ if $z$ is an empty set and is zero otherwise. $A(s|z,\pmb{r}_l^{\rho})$, is equivalent to $A_1(s|r,\pmb{r}_l^{\rho})$ if $z=r$ and equivalent to $A_2(s|\pmb{r}_l^{\rho})$ if $z=.$.
The optimal growth rate is derived by optimizing Eq. (\ref{het6main}) with respect to $A$, subject to the condition $\sum_{s}A(s|z,\pmb{r}^{\rho}_l)=1$. Doing so, we can formally write for the optimal growth rate (Appendix \Aplocalopt):
\begin{align}
\Lambda_{(R,G,C)}^{l*}=\sum_{\epsilon=1}^{n_{\epsilon}}p_{\epsilon}\log w_{\epsilon}-H^{\pmb{N}_q\pmb{N}_l}_{p_{\epsilon}}(\epsilon|z,\pmb{r}^{\rho}).
\label{decay6main}
\end{align}
Where we have defined $H^{\pmb{N}_q\pmb{N}_l}_{p_{\epsilon}}=\min_{A}-\sum_{\epsilon}p_{\epsilon}\log(A\pmb{N}_q\pmb{N}_l(s_{\epsilon}|\epsilon))$. This has been introduced as a generalization of the concept of entropy \cite{Rivoire}. Here, it can be thought of as quantifying the remaining environmental uncertainty given its description by the CSS.
By subtracting the optimal growth of a similar population who lacks the CSS in Eq. (\ref{lambda1}) from Eq. (\ref{decay6main}), we derive for the value of the CSS:
\begin{align}
V_{(R,G,C)}^l=\Lambda_{(R,G,C)}^{l*}-\Lambda_0^*=I^{\pmb{N}_q\pmb{N}_l}_{p_{\epsilon}}(\epsilon;z,\pmb{r}^{\rho}).
\label{decay7main}
\end{align}
Where we have defined  $I^{\pmb{N}_q\pmb{N}_l}_{p_{\epsilon}}(\epsilon;z,\pmb{r}^{\rho})=H_{p_{\epsilon}}-H_{p_{\epsilon}}^{\pmb{N}_q\pmb{N}_l}$. This has been introduced as a generalizations of mutual information \cite{Rivoire}. Here, it can be thought of as the information content of the CSS from the world.\\
Expressions in Eq. (\ref{decay6main}) and Eq. (\ref{decay7main}) cannot be made any simpler. However, as shown in Appendix \Aplocalheto, using concavity of the logarithm, it is possible to derive the following upper and lower bounds for them in terms of Shannon entropies and mutual information:
\begin{align}
&0 \leq H^{\pmb{N}_q*\pmb{N}_l}_{{p_{\epsilon}}}(\epsilon|r,\pmb{r}^{\rho})\leq q H_{R\pmb{C}_l*\pmb{N}_l}(\epsilon|r,\pmb{r}^{\rho})\nonumber\\&+(1-q)H_{\pmb{C}_l*\pmb{N}_l}(\epsilon|\pmb{r}^{\rho}),\nonumber\\&H_{p_{\epsilon}}\geq V_{(A,G,C)}^l \geq q I_{R\pmb{C}_l*\pmb{N}_l}(\epsilon;r,\pmb{r}^{\rho})+(1-q) I_{\pmb{C}_l*\pmb{N}_l}(\epsilon;\pmb{r}^{\rho}).
\label{ineqhet}
\end{align}
Here, we have made convolution signs specific to avoid ambiguity. In Appendix \Aplocalheto, we give an interpretation to these inequalities in terms of positive effects of various sources of heterogeneity. In fact, loosely speaking, the right hand side of the second inequality in Eq. (\ref{ineqhet}) can be thought of as the value of acquired information for a population whose individuals use the same channels as used in the CSS, to acquire information from environmental conditions, but without having a CSS. Inequality Eq. (\ref{ineqhet}) shows then, how a CSS, by employing a distributed information acquisition system, amplifies the effect of acquired information on growth using heterogeneity in communication.\\
To see the positive effect of heterogeneity in a more manifest way, we consider a population in which a fraction $q$ of the population detect the world state directly and others do not. In addition, all the population receive a profile of signals $\pmb{\sigma}$ drawn according to the distribution $\pmb{N}_{l}(\pmb{\sigma}|\epsilon)$ homogeneously, from a set of individuals who have observed the world. It is easy to see that the population evolution equation in this case is $N_{t+1}= \sum_{\pmb{r}^{\rho}} w_{\epsilon_{t}}(\sum_rqA(s_{\epsilon_t}|r,\pmb{r}^{\rho})R(r|\epsilon_t)+(1-q)A(s_{\epsilon_t}|\pmb{r}^{\rho}))\pmb{C}(\pmb{r}^{\rho}|\pmb{\sigma}_t)N_t$. Iterating this equation from time $0$ to $T-1$, dividing by initial population size, taking the logarithm and dividing by $T$, using stationarity to replace time summation with a summation over states, we derive for the growth rate (Appendix \Aplocalhet):
\begin{align}
\Lambda^{'}=\sum_{\epsilon,\pmb{\sigma}}\pmb{N}_l(\pmb{\sigma}|\epsilon)p_{\epsilon}\log(w_{\epsilon}A\pmb{N}_q(s_{\epsilon}|\epsilon,\pmb{\sigma})).
\label{he5main}
\end{align}
Using the concavity of the logarithm, it is easy to see $\Lambda'\leq\Lambda_{(R,G,C)}^l$ (Appendix \Aplocalhet).
Thus, we see that the value of CSS in the presence of heterogeneity such that each individual receives a signal profile $\pmb{\sigma}$ in each environmental state $\epsilon$ according to the distribution $\pmb{N}_{l}(\pmb{\sigma}|\epsilon)$ is greater than a case when a profile $\pmb{\sigma}$ is drawn according to the same distribution $\pmb{N}_{l}(\pmb{\sigma}|\epsilon)$ at once and transmitted by a subset of individuals to the whole population homogeneously. The reason is that in the former heterogeneous case, heterogeneity enhances the growth by providing diversity and making the population well prepared for errors in signal formation and transmission. In Appendix \Aplocalhet, we show that this result is still valid in a more general growth model with non-diagonal multiplication rates, $w_{\epsilon,s}\neq0$ for $s \neq s_{\epsilon}$.\\
\subsection{local communication: computational studies}
\begin{figure*}
	\centering
	\includegraphics[width=1\linewidth]{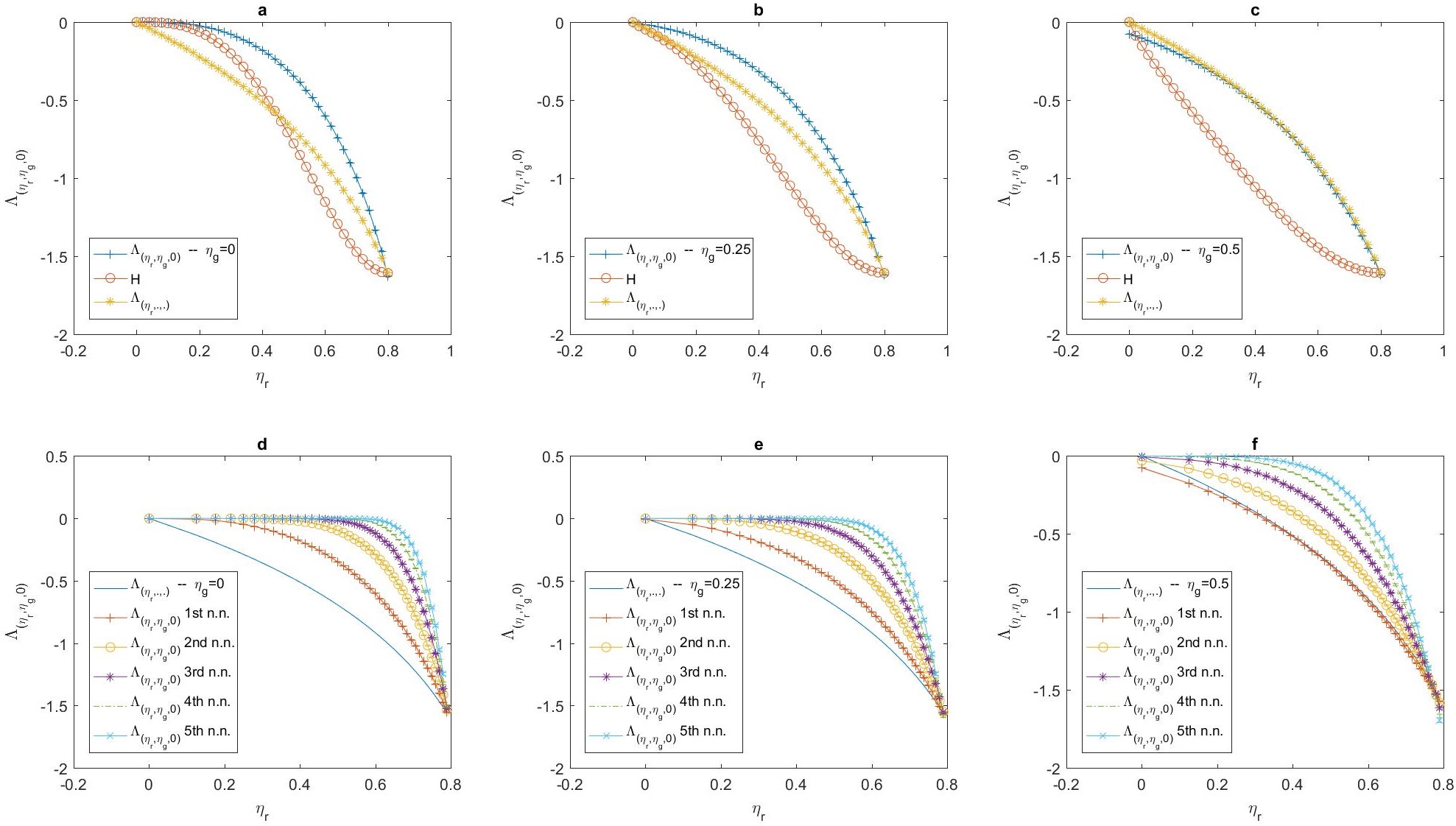}
	\caption{(a), (b), and (c): Growth rate on a square lattice with nearest neighbour interactions for a population using a majority rule as its decision making strategy, versus noise in representation $\eta_r$, for different levels of noise in production. The lower bound derived in the text, $H$, and the growth rate of a population with the same representation channel but without communication is plotted as well. Here $q=1$. As can be seen in (c), by increasing noise in communication, the value of communication decreases. (d), (e), and (f): Growth rate on a square lattice for up to $l$th nearest neighbour interactions as a function of noise in representation, for different values of noise in production is plotted. Here $q=1$. The growth rate approaches maximum growth exponentially as the noise in representation decreases near maximum noise level. Besides, increasing the extent of communications increases growth such that the population can remove all the environmental uncertainty even for high representation and production noise regime.}
	\label{figmainRep}
\end{figure*}
\begin{figure*}
	\includegraphics[width=1\linewidth]{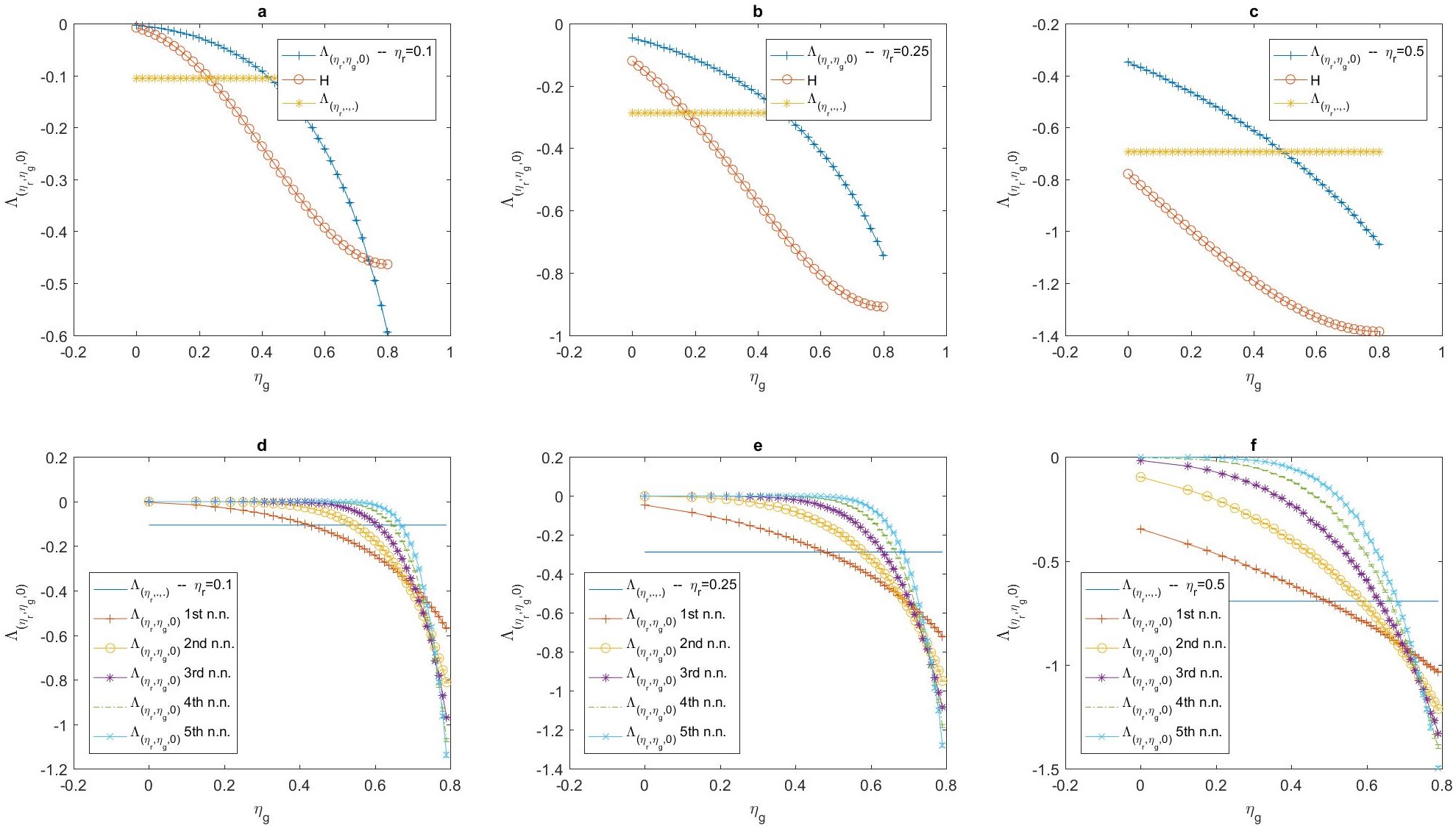}
	\caption{(a), (b), and (c): Growth rate on a square lattice with nearest neighbour interactions for a population using a majority rule as its decision making strategy, versus noise in production $\eta_g$, for different levels of noise in representation. The lower bound derived in the text, $H$, and the growth rate of a population with the same representation channel but without communication is plotted as well. Here $q=1$. As can be seen in the plots, by increasing noise in communication, its value decreases. (d), (e), and (f): Growth rate on a square lattice for up to $l$th nearest neighbour interactions as a function of noise in production, for different levels of noise in representation is plotted. Here $q=1$. The growth rate approaches maximum growth exponentially as the noise in production decreases near maximum noise level. Besides, increasing the extent of communications increases growth such that the population can remove all the environmental uncertainty even for high levels of noise in representation and production.}
	\label{figmainPer}
\end{figure*}
\begin{figure*}
	\centering
	\includegraphics[width=1\linewidth]{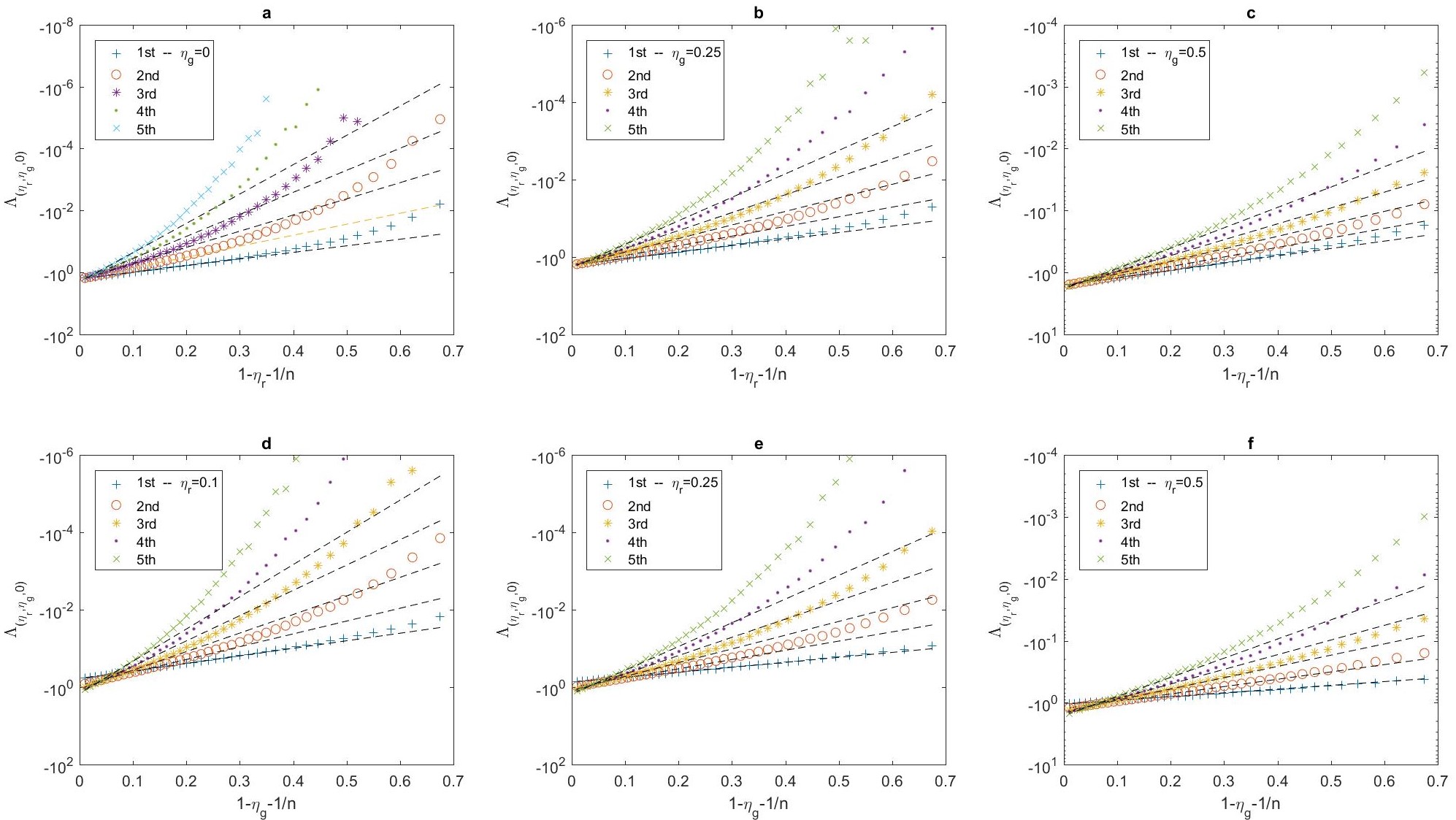}
	\caption{(a), (b), and (c): Log linear plot of the growth rate versus $1-\eta_r-\frac{1}{n}$ for different levels of noise in production. This is the probability by which a correct representation occurs above a uniformly random representation formation. We see that the growth rate increases exponentially as the probability of correct representation formation increases beyond a uniformly random representation formation matrix. Here $q=1$. The slope is non-universal and depends on the extent of communications, $\eta_g$, and $q$. (d), (e), and (f): Log linear plot of the growth rate versus $1-\eta_g-\frac{1}{n}$ for different levels of noise in representation. This is the probability by which a correct signal is produced above a uniformly random signal production. We see that the growth rate increases exponentially as the probability of correct signal production increases beyond a uniformly random signal production matrix. Here $q=1$.}
	\label{figmainexp}
\end{figure*}
\begin{figure*}
	\centering
	\includegraphics[width=1\linewidth]{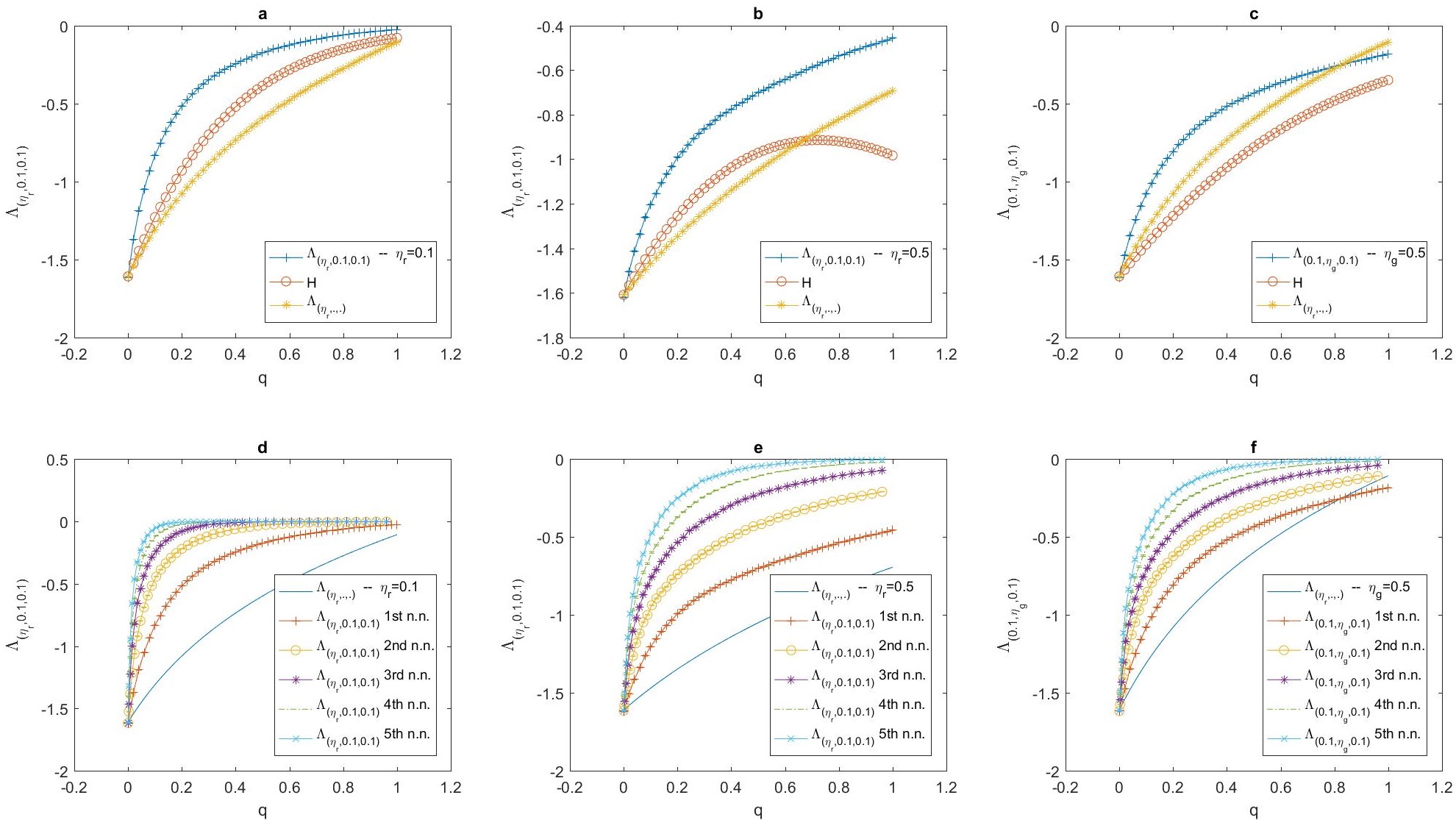}
	\caption{(a), (b), and (c): Growth rate on a square lattice with nearest neighbour interactions for a population using a majority rule as its decision making strategy versus observation probability $q$, for different levels of noise in production and representation. The lower bound derived in the text, $H$, and the growth rate of a population with the same representation channel but without communication is plotted as well. We see that communication increases growth for all observation probabilities $q$, and its value decreases with increasing noise in communication. When noise in communication (here, noise in production) is high, as in (c), with a majority rule the growth rate falls below that of a population without communication. (c), (d), and (e): growth rate on a square lattice for up to $l$th nearest neighbour interactions versus $q$. We see that as the extent of communication increases, the growth rate increases, such that the population can remove all the environmental uncertainty with a simple majority rule, even when the individual observation probability is low, or noise in representation and production is high.}
	\label{figqmain}
\end{figure*}
For the computational experiments for the model with local communication, presented in this section, we consider a CSS with equal number of representations, signals, strategies and environmental states, $n$. As we want to characterize the effect of noise, we parametrize the CSS matrices with a single parameter, the probability of error, or noise $\eta$, such that the diagonal elements of the matrices are $1-\eta$ and all the off diagonal elements are $\frac{\eta}{n-1}$. That is, for example $R(r_i|\epsilon_j)=1-\eta_r$ if $r_i$ is the representation corresponding to environmental state $\epsilon_j$ (if $i=j$), and equals $\frac{1}{\eta}$, if $r_i$ does not correspond to $\epsilon_j$ ($i\neq j$). With this parametrization, the model has $5$ parameters, noise in representation $\eta_r$, noise in production $\eta_g$, noise in comprehension $\eta_c$, observation probability $q$, and the number of states $n$. Among these parameters, in a model of local communication with infinite population, noise in production and comprehension have generally the same effect. This can be seen in Eq. (\ref{het6main}), in which the combination $\pmb{C}\pmb{G}$ appears in the growth rate which is commutative for our parametrization of the CSS matrices. This means that we can interchange $C$ and $G$, which implies, the effect of an amount of noise in comprehension and production is the same. Taking advantage of this fact, we investigate the behaviour of growth rate as a function of the remaining set of parameters. In the following, we will generally subtract the constant term $\langle\log w_{\epsilon}\rangle$, from the growth rate and plot the remaining entropic term. Here the angle brackets denote an average over states $\langle\log w_{\epsilon}\rangle=\sum_{\epsilon}p_{\epsilon}\log \epsilon$. For the communication network we consider a square lattice. This can be considered as a discretized approximation of a homogeneous $2$ dimensional spatial structure as well. We consider the case where individuals transmit their signals to all the neighbours closer than or equal to $l$th neighbour. To investigate different signal decay lengths, we set $l=1$ to $l=5$. The growth rate depends on the decision making strategy individuals use. The optimization problem leading to Eq. (\ref{decay6main}), is not solvable in general. Besides, the optimal action matrix can be very complicated and far from realistic. Instead, we are more interested in the behaviour of growth rate using more plausible, simple and practical decision making rules. A reasonable and simple rule, which we will consider here, is a majority rule. According to this decision making rule, individuals choose the strategy $s$ which is suggested by the majority of their internal representations. That is, individuals count their internal representations and choose the strategy which corresponds to the representation which has occurred with highest frequency in their internal state. This can be thought of a discrete version of a simple averaging rule which is known to be the chi square optimal strategy \cite{Cover}. If there are more than one representation which occurs the highest number of times, the individual randomly chooses the strategy corresponding to one of her majority representations. Unless one of the majority representations is supported by her personal observation, in which case the individual chooses the strategy corresponding to this representation. We will see that using this simple decision making rule the population can come very close to the optimal growth provided noise is not too high. In the figures, $\Lambda_{(\eta_r,\eta_g,\eta_c)}$, means the growth rate of a population whose representation, production, and comprehension matrices are given, respectively by the noise parameters, $\eta_r$, $\eta_g$, and $\eta_c$. For details of simulations see Appendix \Apsim.\\
To show how the growth rate behaves, we begin by considering a first nearest neighbour communication $l=1$, on a square lattice. In Fig. (\ref{figmainRep}.a) to Fig. (\ref{figmainRep}.c) we plot the growth rate as a function of noise in representation $\eta_r$, fixing the observation probability $q=1$, for different values of noise in production $\eta_g$, setting $\eta_c=0$ and $n=5$. We also plot the lower bound on optimal growth $H=-(q H_{R\pmb{C}_l*\pmb{N}_l}(\epsilon|r,\pmb{r}^{\rho})+(1-q)H_{\pmb{C}_l*\pmb{N}_l}(\epsilon|\pmb{r}^{\rho}))$ derived in Eq. (\ref{ineqhet}), denoted as $H$ in the figures. Furthermore, to see how much the growth rate increases solely due to communication, we calculate the growth rate of the same population, with the same observation probability $q$, representation matrix $R(r|\epsilon)$ and decision making rule or action matrix, but without communication, and denote it by $\Lambda_{(R,.,.)}$ in the figures. The difference between growth rate and $\Lambda_{(R,.,.)}$ can be thought of as a measure of how much the growth rate is increased solely due to communication. We will occasionally refer to it as the value of communication and denote it by $\Lambda_{(\eta_r,\eta_g,\eta_c)}^c$.\\
First, we note that for low noise levels in communication, communication increases growth, even when all the individuals observe the world. This can be seen in the figures as the growth rate with communication lies above that in the absence of communication $\Lambda_{(R,.,.)}$. The reason is that communication enables individuals to reach higher information from the environment, by effective repetitive observations through information sharing. As noise in communication increases, the signals received through communication become misleading, and communication can decreases the growth rate such that it falls below $\Lambda_{(R,.,.)}$, as can be seen in Fig. (\ref{figmainRep}.c). This can be seen more clearly in Fig. (\ref{figmainPer}.a) to Fig. (\ref{figmainPer}.c), where the growth rate as a function of $\eta_g$, for different values of $\eta_r$, fixing $q=1$, $\eta_c=0$, and $n=5$ is plotted. We see that as noise in communication increases, the growth rate falls below that of a population without communication. That a highly noisy communication decreases growth, is a consequence of a majority rule strategy. As signals received through a highly noisy communication become misleading, relying on them in decision making, leaves the population worse off, compared to when not communicating. The population can cure this by changing its decision making rule. For example, a better rule in such high noise level is to ignore communication altogether, in which case the growth rate can not fall below that of a population without communication. Another simple rule which performs better in high communication noise level, is that individuals rely on communication in decision making, only when not making a personal observation. It is obvious that such a rule can not decrease growth below the growth rate of a population without communication. However, such rules perform worse in low noise levels. This can be seen in Fig. (\ref{figmainPer}.a) to Fig. (\ref{figmainPer}.c), as relying on communication increases growth above $\Lambda_{(R,.,.)}$ in low noise levels. That in a given population which decision making rule performs better, depends on noise level. Given the noise level in the CSS, evolution can select one of these simple rules.\\
We turn to the lower bound on growth derived in Eq. (\ref{ineqhet}) by noting that, as here the growth rate is calculated for a majority rule decision making strategy, which is not the optimal strategy, it is not necessary that it respects this lower bound. However, as can be seen in Fig. (\ref{figmainRep}.a) and Fig. (\ref{figmainRep}.b), the lower bound is respected with a majority rule strategy for low values of noise in communication. The same conclusion is visible in Fig. (\ref{figmainPer}.a), where we see that for very high $\eta_g$, the lower bound can be violated with a majority rule strategy. However, the region where this lower bound is not respected corresponds to the region where communication is detrimental. As discussed, biologically this region should be pruned away by evolution by selecting a better decision making strategy. We see in Fig. (\ref{figmainRep}.a) to Fig. (\ref{figmainRep}.c), and in Fig. (\ref{figmainPer}.a) to Fig. (\ref{figmainPer}.c), that for all the region where communication is beneficial the lower bound is respected with a simple majority rule strategy.\\
In Fig. (\ref{figmainRep}.d) to Fig. (\ref{figmainRep}.f), we investigate the effect of increasing the extent of communications. We do this by considering a population residing on a square lattice in which individuals transmit their signals to their up to $l$th nearest neighbours. The growth rate for the cases when $l=1$, to $l=5$ versus $\eta_r$, for different values of $\eta_g$, setting $q=1$, $\eta_c=0$, and $n=5$ is plotted. We see that as the extent of communications increases, the growth rate approaches maximum possible growth in which the population can extract all the information available in its CSS from environment, even for relatively large values of noise in representation $\eta_r$ and production $\eta_g$. Here, we have denoted the growth of the same population without communication $\Lambda_{(R,.,.)}$, and can be seen that the increase in growth solely resulted from communication, generally increases as the extent of communication increases.\\
To investigate the effect of noise in production, in Fig. (\ref{figmainPer}.d) to Fig.(\ref{figmainPer}.f), we do a similar experiment by plotting the growth rate as a function of $\eta_g$ for different values of $\eta_r$, fixing $q=1$, $\eta_c=0$, and $n=5$. The same conclusion is derived here by noting that the growth rate approaches maximum growth for relatively high values of communication noise when the extent of communications increases. By comparing to similar plots in Fig. (\ref{figmainRep}), we conclude that the detrimental effect of an amount of noise in representation is generally higher than the same noise in communication. This is so as noise in representation also impacts the information individuals can draw relying solely on their own personal observation, while noise in production only deteriorates the information derivable from communication.\\
We turn to an interesting question, how the growth rate behaves as the noise in CSS matrices decreases? Evolutionarily, this question is an important question, as by quantification of the growth rate as a function of noise reduction in CSS matrices, we are able to asses if the fitness effect of noise reduction is high enough to explain the emergence of a CSS. More precisely, we ask, starting with uniformly random CSS matrices, which gives any of the possible outcomes for a state with the same probability, how the reduction of some amount of noise, affects the growth? Interestingly, as we will shortly see, the growth rate increases exponentially as noise in CSS matrices is reduced, near the maximum noise level. This implies noise reduction has high fitness consequence and thus, the emergence of CSS in an evolutionary dynamics, with small noise reduction steps should be a plausible and easy outcome. This can explain the prevalence of such CSSs in the biological world.\\
In Fig. (\ref{figmainexp}.a) to Fig. (\ref{figmainexp}.c), we plot the same growth rates plotted in Fig. (\ref{figmainRep}.a) to Fig. (\ref{figmainRep}.c), versus $1-\eta_r-\frac{1}{n}$ in a semi-logarithmic plot to confirm the exponential increase of growth rate as noise in representation decreases. $1-\eta_r$ is the probability of correct representation formation and $\frac{1}{n}$ is the probability that the correct representation occurs with a uniformly random representation formation. Their difference thus, is a measure of how representation formation occurs better than uniformly at random, or equivalently a measure of noise reduction in representation. We see in Fig. (\ref{figmainexp}) that the growth rate increases exponentially as noise in representation is reduced near the maximum noise level (a uniformly random representation formation). Furthermore, the range of exponential increase in growth by noise reduction in rather wide: it holds from maximum noise level (minimum growth), up to a small distance of maximum growth. This can be seen by comparing Fig. (\ref{figmainexp}.a) to Fig. (\ref{figmainexp}.c), with Fig. (\ref{figmainRep}.d) to Fig. (\ref{figmainRep}.f). This exponential increase in growth rate as a function of reduced noise in representation, shows that, by noise reduction in high noise levels, a population can rapidly increase its growth rate. This exponentially positive effect of noise reduction makes the development of a CSS an easy task in an evolutionary process. In Fig. (\ref{figmainexp}), we see that the slope of the fit is non-universal and depends on $\eta_g$. Generally the speed of increase in growth by noise reduction is accelerated by communication and increases by decreasing noise in communication and increasing the extent of communication (see Fig. (\ref{figmainRep}.d) to Fig. (\ref{figmainRep}.f)). This shows that communication amplifies the effect of noise reduction in representation. Evolutionarily this has the consequence that there is higher pressure for noise reduction on biological sensors in populations who communicate, compared to populations who does not communicate. In the computational experiments presented here, we have set $q=1$. In Appendix \Apsimq, we confirm that this exponential behaviour holds also in other values of $q$ and the slope of the fit depends on $q$ for low values of $q$ and saturates to a $q$-independent value for high $q$s.\\
In Fig. (\ref{figmainexp}.d) to Fig. (\ref{figmainexp}.f), we repeat the same procedure for noise in production. Here, we plot the growth rate as a function of $1-\eta_g-\frac{1}{n}$, in a semi-logarithmic plot, and are able to establish the exponential increase in growth as noise in production is reduced near maximum noise level. However, we note that high production noise level, coincides with the region where communication is detrimental and as argued before, this regime is not of practical relevance biologically. Nevertheless, by comparing Fig. (\ref{figmainexp}.d) to Fig. (\ref{figmainexp}.f), with Fig. (\ref{figmainPer}.d) to Fig. (\ref{figmainPer}.f), we see that the domain of exponential increase in growth rate by noise reduction extends to the biologically interesting region where communication is beneficial. This exponential behaviour shows the ease with which communication can be developed in an evolutionary process as the exponential increase of growth rate, makes noise reduction highly adaptive. The slope of the fit in fact depends on other parameters of the model, including the number of states $n$. In appendix \Apsimnc, we see that when $n$ increases, the slope decreases. This implies with high number of states, the benefit of noise reduction is lower near maximum noise level. The reason is that the probability that a uniformly random noise gives rise to a situation where an erroneous representation happens with high probability with a majority rule strategy, decreases with the number of states and thus the detrimental effect of noise decreases.\\
In Fig. (\ref{figqmain}), we turn to the question of the behaviour of the growth rate as a function of observation probability. We plot the growth rate as a function of $q$, for different values of $\eta_r$, and $\eta_g$, setting $\eta_c=0.1$ and $n=5$. By comparing Fig. (\ref{figqmain}.b) and Fig. (\ref{figqmain}.c), we see that, as argued before, an amount of noise in representation generally has more detrimental effect than the same amount of noise in production, for all values of $q$. We saw before, for example in high $\eta_g$ regime in Fig. (\ref{figmainPer}.a) to Fig. (\ref{figmainPer}.c), that highly noisy communication is detrimental in the sense that the population is better off if individuals do not take communication into account if have made a direct observation. However, we had set $q=1$ there. In Fig. (\ref{figqmain}), we can see the performance of noisy communication improves as $q$ decreases. This is so because for lower $q$, fewer individuals have observed the world directly and can possibly become misled by a highly noisy communication, while others who have not made a personal observation can still benefit from even a highly noisy communication in absence of any other cue in decision making.\\
In Fig. (\ref{figqmain}.d) to Fig. (\ref{figqmain}.f), we plot the growth rate as a function of $q$ for different extents of communications $l=1$ to $l=5$, for different values of $\eta_r$ and $\eta_g$, setting $\eta_c=0.1$ and $n=5$. We see that as the extent of communications increase the growth rate approaches maximum growth for relatively low values of $q$. Calling $q_{l,\delta}$ the minimum value of $q$ for which the growth rate of a population having up to $l$th order interactions comes to a small distance $\delta$ of the maximum growth, we see in the figure that $q_{l,\delta}$ generally increases with $l$ and decreases with noise level.\\
We see that generally the growth rate increases rapidly with increasing $q$ near $q=0$, especially when the extent of communications increase. This interesting effect shows how communication enables the population to increases its growth and reaches optimal growth, by amplification of information entered to the population via observation by few individuals.\\
\section{Global communication}
The rapid increase of growth rate by increasing the extent of communications, we saw in the last section, hints us at how the growth rate behaves in the limit that the signal decay length scale approaches infinite length. This is a regime of global communication, in the sense that a signal transmitted by an individual is not restricted to a local neighbourhood and reaches the whole population. We can consider this as a regime with infinite extent of communication. With this interpretation, the fact that by increasing the extent of communication $l$, the population is able to achieve maximum growth even with low observation probability and high noise level in its CSS, hints us that in the global communication regime, the population should be able to extract all the information from environment available in its CSS. We will show this in this section.\\
To work with a specific model we assume that each individual observes the world state with probability $q$ and transmit a signal to a fraction $b$ of the population to inform them of her observation. This can result if each individual as receiver, receives a transmitted signal with probability $b$. In this model, the average number of signals an individual receives is $m=bqN_t$. As the population size is growing (assuming positive growth rate which is equivalent to non-extinction), the number of signals is increasing with time. On the other hand, receiving a signal profile composed of $m$ signals, is effectively the same as making $m$ observations through the channel $CGR(r^{\rho}|\epsilon)$. With high enough $m$ the receiver samples the whole distribution $CGR(r^{\rho}|\epsilon)$. This will happen in a large enough time if $m$ is increasing with population size, as in the model of global communication. The fact that the individuals are able to sample the whole distribution $CGR(r^{\rho}|\epsilon)$, allows us to introduce an effective signal $o$, which designates the conditional distribution $CGR(r^{\rho}|\epsilon)$. That is the effective signal $o$ is different for any two environmental states, $\epsilon_1$ and $\epsilon_2$, if and only if $CGR(r^{\rho}|\epsilon_1)\neq CGR(r^{\rho}|\epsilon_2)$. Consequently, This signal divides the environmental states into classes. The members of each class are indistinguishable from each other by $CGR$ (i.e. for any $\epsilon_1$ and $\epsilon_2$ in a class $CGR(r^{\rho}|\epsilon_1)= CGR(r^{\rho}|\epsilon_2)$) and are distinguishable from other classes (i.e. for any $\epsilon_1$ and $\epsilon_2$ in two different classes $CGR(r^{\rho}|\epsilon_1)\neq CGR(r^{\rho}|\epsilon_2)$). The signal $o$ contains all the information from environment contained in $\pmb{r}^{\rho}$. In fact, individuals can use sub-optimal strategies which depend on the details of $\pmb{r}^{\rho}$. But, as long as the optimal growth and optimal decision making rule is concerned, individuals can neglect the details of the profile $\pmb{r}^{\rho}$, and look at the typical sequence to which $\pmb{r}^{\rho}$ belongs. The signal $o$ gives this typical sequence.  The effective signal $o$ is drawn according to a distribution $O(o|\epsilon)$. This distribution can easily be derived given the CSS matrices. An example of this effective signal and its distribution is given in the Appendix \Apglob.\\
Therefore, as long as the optimal growth is concerned, our model becomes effectively equivalent to a model in which each individual observes the world state using $R(r|\epsilon)$ with probability $q$. In addition, in each environmental state, the whole population receives a signal $o$ according to $O(o|\epsilon)$. The population evolution equation becomes $N_{t+1}=w_{\epsilon}(qA_1(s_{\epsilon}|r,o_t)R(r|\epsilon)+(1-q)A_2(s_{\epsilon}|o_t))N_t$. Iterating this equation from time zero to $T-1$, dividing by initial population size, taking the logarithm, dividing by $T$ and taking the large time limit, and using the stationarity of the process to replace time summation with a summation over states, we have for the long term growth rate
$\Lambda_{(R,G,C)}^g=\sum_{\epsilon,o}O(o|\epsilon)p_{\epsilon}\log(w_{\epsilon}(\sum_{r}qA_1(s_{\epsilon}|r,o)R(r|\epsilon)+(1-q)A_2(s_{\epsilon}|o)))$.
As usual, we define some abbreviated notation. We define a new random varible $z$ which can take eighther an empty set $.$ or $r$. In addition we define a new action matrix $A(z,o)$ such that $A(s|z=r,o)=A_1(s|r,o)$ and $A(s|z=.,o)=A_2(s|o)$. We also generalize the definition of representation matrix such that $R(z=r|\epsilon)=R(r|\epsilon)$ and $R(z=.|\epsilon)=0$. Furthermore, we define a new probability distribution $\pmb{B}(z|\epsilon)=qR(z|\epsilon)+\delta_{z,.}(1-q)$. With these definitions, the growth rate can be written:
\begin{align}
\Lambda_{(R,G,C)}^{g}=&\sum_{\epsilon,o}O(o|\epsilon)p_{\epsilon}\log(w_{\epsilon}A\pmb{B}(s_{\epsilon}|o,\epsilon)).
\label{nons002}
\end{align}
The optimal growth rate is derived by optimizing this with respect to $A$ subject to the constraint $\sum_{s}A(s|z,o)=1$. Doing so, we derive for the optimal growth rate:
\begin{align}
\Lambda_{(R,G,C)}^{g*}=\sum_{\epsilon}p_{\epsilon} \log w_{\epsilon} -H^{\pmb{B}}_{Op_{\epsilon}}(\epsilon|z,o).
\label{nons003}
\end{align}
Where we have defined: $H^{\pmb{B}}_{Op_{\epsilon}}(\epsilon|z,o)=\max_{A}\sum_{\epsilon,o}O(o|\epsilon)p_{\epsilon}\log(A\pmb{B}(s_{\epsilon}|o,\epsilon))$.
The value of the CSS is derived by subtracting the growth rate of a population without a CSS in Eq. (\ref{lambda1}) from Eq. (\ref{nons003}):
\begin{align}
V_{(R,G,C)}^{g}=H_{p_{\epsilon}}(\epsilon)-H^{\pmb{B}}_{Op_{\epsilon}}(\epsilon|z,o)=I^{\pmb{B}}_{Op_{\epsilon}}(\epsilon;z,o).
\label{nons00003}
\end{align}
Making the strategy conditional on the representation resulted from direct observation is useful only when two conditions are satisfied. First, there are some environmental states which are indistinguishable using $CGR$. Second, $R$ can distinguish at least two environmental states that are not distinguishable by $CGR$. That is there are at least two environmental states $\epsilon_1$ and $\epsilon_2$, for which $R(r|\epsilon_1)\neq R(r|\epsilon_2)$ and $CGR(r|\epsilon_1)= CGR(r|\epsilon_2)$. When there are no states that are indistinguishable using $CGR$ but distinguishable using $R$, $r$ provides no new information and can be dropped out of $H^{\pmb{B}}_{Op_{\epsilon}}$. In this case the optimal growth rate simplifies  to $\Lambda_{(R,G,C)}^{g*}=\max_{A}\sum_{\epsilon,o}O(o|\epsilon)p_{\epsilon}\log(w_{\epsilon}(A(s_{\epsilon}|o)))$, which, by performing optimization becomes $\Lambda_{(R,G,C)}^{g*}=\langle \log w_{\epsilon} \rangle -H_{Op_{\epsilon}}(\epsilon|o)$. The value of the CSS becomes:
\begin{align} V_{(R,G,C)}^{g}=H_{p_{\epsilon}}(\epsilon)-H_{Op_{\epsilon}}(\epsilon|o)=I_{Op_{\epsilon}}(\epsilon;o).
\label{mutual}
\end{align}
Here, $H_{Op_{\epsilon}}(\epsilon|o)=-\sum_{\epsilon,o}O(o|\epsilon)p_{\epsilon}\log O(o|\epsilon)$ and $I_{Op_{\epsilon}}(\epsilon;o)=H_{p_{\epsilon}}(\epsilon)-H_{Op_{\epsilon}}(\epsilon|o)$ are standard Shannon (conditional) entropy and mutual information.\\
Another case of practical interest is when all the environmental states are distinguishable in $CGR$. In this case, $o$ becomes a perfect signal which identifies the environmental state. In this case we have $H_{Op_{\epsilon}}(\epsilon|o)=0$ and $I_{Op_{\epsilon}}(\epsilon;o)=H_{p_{\epsilon}}(\epsilon)$. Thus the value of the CSS becomes equal to the environmental entropy as the CSS can remove all the environmental uncertainty.
\begin{align}
V_{(R,G,C)}^{ns.g}=H_{p_{\epsilon}}(\epsilon).
\label{nons06}
\end{align}
Intuitively, these information theoretic terms are also information content of such a language-like system from the world. The appearance of entropy and mutual information in the effect of a CSS on growth, further encourages its interpretation as the information content of the CSS. Based on this intuition, we can argue that the information content of a language-like system, such as CSS from the world, is given by the same expressions we have derived as their value. That is Eq. (\ref{decay7main}), Eq. (\ref{nons00003}), Eq. (\ref{mutual}), and Eq.(\ref{nons06}), depending on the structure of communications. Finally, we note that although we have derived theses terms for the value of a CSS in a model with global communication, our discussion in the last section shows that the value of CSS in a model with local communication approaches these terms as well, as the extent of communications increase.
\subsection{global communication: computational studies}
\begin{figure}
	\includegraphics[width=1\linewidth]{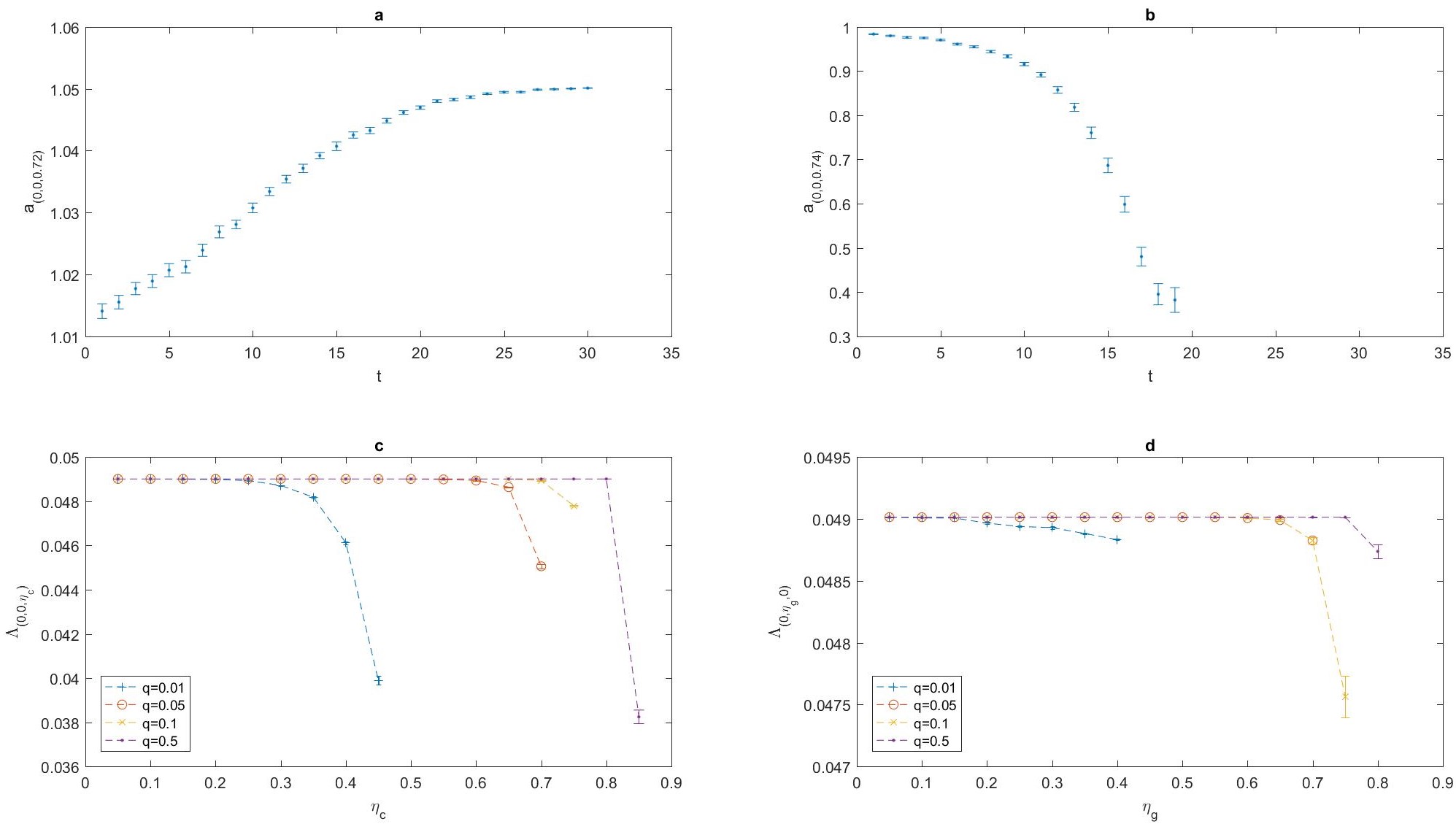}
	\caption{Global communication. (a) and (b): The instantaneous growth rate $a_t$ for global communication as a function of time $t$. We see that as long as the initial population size is above extinction threshold, the population has accelerated growth. While in (b) with a slight increase in $\eta_c$ the population goes to extinction with a decelerating rate. Here $N_0=1000$, $\eta_r=0$, $\eta_g=0$, and $n=10$. Averages and standard errors are calculated based on a sample of $R=30$ realizations. (c): Growth rate versus $\eta_c$ for different fraction of observing individuals $q$. The growth rate is calculated in a time window of $T=30$ time steps, and an average over $R=10$ realizations is taken. We see that for each $q$, below a critical $\eta_c$ the population has positive growth. Due to increase in growth by increasing population size, the growth rate in any region with positive growth will saturate to optimal growth, in long time limit. Here $N_0=1000$, $\eta_r=0$ and $\eta_g=0$, and $n=10$. (d): growth rate versus $\eta_g$ keeping the other simulation parameters the same as in (c).}
	\label{figglobal}
\end{figure}

We start this section, by characterizing the behaviour of the instantaneous growth. We define the instantaneous growth as the fractional increase in population in one time step:
\begin{align}
a_{t}=\frac{N_{t+1}}{N_{t}}.
\label{instan}
\end{align}
Denoting by $d_t$, the fraction of the population who choose the correct strategy $d_t=\frac{N_{t,{\epsilon}_t}}{N_t}$, by using Eq. (\ref{popev}), we have $a_t=w_{\epsilon_t}d_t$. We argued that in a model with global communication, the average number of signals an individual receives, is a function of population size and increases with population growth. This implies that, in larger populations, individuals have higher information from environmental conditions, and are able to infer the environment with higher accuracy. That is $d_t$ and thus $a_t$ increase with population growth. This means that population growth is accelerating in a growing population. The growth rate saturates to the optimal value when the population becomes large enough so that the individuals are able to extract all the information from environment available in the CSS.\\
To confirm accelerating growth, we appeal to a computational experiment. We consider a population evolving in an environment with $n=10$ states, with a uniform environmental probability distribution. We choose the multiplication rate to be a diagonal matrix with all the elements equal to $w_{\epsilon}=1.05$, and parametrize the CSS matrices with a single noise parameter as before. We choose $\eta_r=\eta_g=0$. Individuals use the same simple majority rule as before, as their strategy. We start with an initial population of size $N_0=1000$. In each environmental state, a fraction $q=0.05$ of the individuals observe the world and transmit a signal to others. The averages and standard errors are calculated based on a sample of $R=50$ simulations. In Fig. (\ref{figglobal}.a), we plot $a_{(0,0,0.72)}$ as a function of time, where we can see that the population has accelerating growth. The growth rate saturates to a constant value equal to $a_t=1.05$, in large times. This is when the population becomes large enough so that the individuals are able to infer the environmental state with high accuracy.\\
The fact that the population is able to reach optimal growth by a simple majority rule, instead of a complicated best strategy, shows another effect, simplification of decision making due to the rich information provided by the CSS. Information sharing through communication, amplifies information entered to the population through observation by a few individuals. This amplification of information by communication results in simplification of decision making. We saw that the same effect is present in a model of local communication, when the extent of communications increase.\\
As noise in CSS increases, the inference capability of the individuals decreases and so the instantaneous growth rate. If the constant term $\langle \log w_{\epsilon} \rangle$, is low enough, this can lead to an instantaneous growth smaller than $1$, which will lead to population extinction given enough time. We can see this phenomena in Fig. (\ref{figglobal}.b). Here, we increase the noise by a small amount, and plot $a_{(0,0,0.74)}$. We see that here the instantaneous growth rate becomes smaller than $1$ and decreases with a decelerating rate to such a point where the population goes extinct. Deceleration of instantaneous growth, results from the fact that, as the instantaneous growth is an increasing function of the population size, when the population decreases, the inference capability of individuals, and consequently the instantaneous growth rate, decreases as well.\\
In Fig. (\ref{figglobal}.c) and Fig. (\ref{figglobal}.d) we return to long term growth rate and appeal to computational experiments to confirm the results of the last sub-section. Besides, we want to asses how the growth rate behaves in parameter regimes close to extinction. We consider a population living in an environment with $n=10$ states. The environments are equi-probable. The CSS matrices are given as before, parametrized with a single noise parameter. For any noise level above a uniformly random CSS ($\eta \neq \frac{n-1}{n}$), all the environmental states are distinguishable in this CSS. We choose the multiplication rate to be a diagonal matrix with all the elements equal to $w_{\epsilon}=1.05$. With this choice the optimal growth rate is $\Lambda_{(R,G,C)}*=\langle \log w_{\epsilon} \rangle=\log 1.05=0.0488$, and the value of the CSS is equal to the environmental entropy $H_{p_{\epsilon}}(\epsilon)=2.3026$. In our simulations we start with an initial population of size $N_0=1000$, and consider a case where a fraction $q$ of the individuals observe the world and transmit a signal to the rest of the population. We calculate the growth rate over a window of $T=30$ time steps, defined as $\Lambda=\frac{1}{T}\log(\frac{N_T}{N_0})$. The long term growth rate is derived from this by taking infinite time limit. The reported averages and standard errors are calculated based on a sample of $R=10$ realizations. In Fig. (\ref{figglobal}.a), we plot the growth rate $\Lambda_{(0,0,\eta_c)}$ as a function of $\eta_c$ setting $\eta_r=\eta_g=0$ for different values of $q$. And, in Fig. (\ref{figglobal}.b), we plot the growth rate $\Lambda_{(0,\eta_g,0)}$ as a function of $\eta_g$, setting $\eta_r=\eta_c=0$, for different values of $q$.
The number of signals an individual receive is equal to $qN_0$. Fixing $N_0=1000$, for low levels of noise this number of signals is enough for the individuals to infer the environmental state with accuracy. Consequently, the population will have an instantaneous growth larger than $1$ and increases. Once increasing, the population will obtain higher inference capability such that it can extract all the information from environment available in its CSS, and its long term growth approaches to $\sum_{\epsilon}p_{\epsilon}\log w_{\epsilon}$ in long times. This can be seen in the figures for low values of noise for different $q$s. However, as noise level increases, fixing the initial population size and observation probability, and thus the average number of signals that individuals receive, individuals are not able to infer environmental state with high accuracy, and the population will start with a instantaneous growth smaller than $1$. In long time, such a population will go extinct.\\
\section{discussion}
We quantified the effect of a CSS on the growth rate by information theoretic terms, which can be interpreted as the information content of such language-like system from environment. We considered local communication, and showed that heterogeneity resulted from local communication increases growth by producing diversity. We showed that the effect of a CSS on growth increases exponentially by noise reduction. Considering the evolvability of a CSS in an evolutionary context, this implies that the emergence of a CSS is a plausible and easy evolutionary outcome, as small noise reduction steps have exponential fitness consequences. We also showed that communication amplifies the effect of noise reduction in representation formation. This implies that there should be higher pressure for noise reduction in sensory representation formation in populations who are able to communicate compared to those who are not. Besides, we showed that an amount of noise in representation formation has more disadvantageous effect compared to the same amount of noise in signal production, in a CSS.\\
We argued that a CSS, by endowing the possibility of multiple effective observations, amplifies information entered to the community through observation by few individuals. This advantage can lead evolution to favour a distributed decision making design over a central one, by development of a language like system. Furthermore, the richness of the information an individual receives, hugely simplifies the optimization problem faced by the individuals. Consequently, the population can come close to the optimal growth by 'simple good strategies', such as a majority rule, instead of the 'complicated best strategy'. This removes any need on the part of individuals to a high memory or high computational capacity. Another interesting consequence of information amplification by communication, is accelerated growth. When the community has access to global communication, by population growth, the number of effective observations of each individual increases, and the community as a whole reaches better inference capability. This leads to accelerated growth as the instantaneous growth rate becomes an on average increasing function of population size. Evidence of increase in inference capability of the community by population growth due to communication has been noted. For example in hone bee populations \cite{donald}. Furthermore, empirical evidence for the accelerated growth resulted from population growth has been noted, for example in the case of human population in economic literature \cite{Kremer}.\\
The phenomenon of collective sensing has similarities with language. One can argue that human language can be seen as a highly evolved version of a collective sensing system in which individuals reach richer information of the world by sharing their experiences. In fact, looking at language as a Saussurean system of signs, a similar model to the one introduced here has been proposed as a model of a proto-language \cite{Martin,Komarova,Joshua}. However, that model ignores the relationship of language with outside world and its functional role in advising behaviour and decision making. The model introduced here instead, by including these fundamental aspects of language, can be seen as a simplified model of language seen as a Saussurean system of signs as well. In this perspective, our results can be seen as answers to fundamental questions regarding information content and fitness value of information provided by a language.\\
\section*{acknowledgement} We are indebted to Yasser Roudi for many useful discussions and insightful suggestions, and for his generous helps in conducting the research and preparing the manuscript. M.S. is grateful to Kavli Institute for Systems Neuroscience and Centre for Neural Computation for hospitality during part of this research.


\pagebreak

\setcounter{figure}{0}
\setcounter{equation}{0}
\setcounter{section}{0}

\newcommand\Rivoire{20}
\newcommand\lambd{4}
\newcommand\detmain{6}
\newcommand\detmai{7}
\newcommand\hetmain{5}
\newcommand\decay{9}
\newcommand\ineqhet{8}
\newcommand\hemain{11}
\newcommand\Lyap{2}
\newcommand\secloc{III.B}
\newcommand\eqlyap{2}
\newcommand{\eqhetsixmain}{5}


	\renewcommand\thesection{\Alph{section}}
	\renewcommand\thesubsection{\thesection.\arabic{subsection}}

	\renewcommand\thefigure{A.\arabic{figure}} 
	\renewcommand\theequation{A.\arabic{equation}} 
	\makeatletter
	\def\p@subsection{}
	\makeatother
	
	\clearpage
	\appendix
	
	\begin{center}
		{\large \bf Appendices\\ }
		\vspace{0.4cm}
		
		

	\end{center}
	\vspace{0.4cm}
	\setcounter{page}{1}
	\section{Notation}
	\label{notation}
	In this text bold lower case letters such as $\pmb{x}$ mean sets of elements $x$. We call them a profile of elements $x$. $\sum_{\pmb{x}}'$ means a summation over all possible sizes of the set $\pmb{x}$. $\sum_{x\in\pmb{x}}$ means a summation over all possible values for the members of the set $\pmb{x}$ fixing the number of members of the set. $\sum_{\pmb{x}}$ means a summation over all possible sets $\pmb{x}$, that is all possible set size and all possible value of each member $\sum_{\pmb{x}}= \sum_{\pmb{x}}'\sum_{x\in\pmb{x}}$. Sometimes, for ease, we use $\sum_{\pmb{x}}$ for $\sum_{x\in\pmb{x}}$. If so this should be clear from the context and we will explain it. Bold upper case letters denote transition matrices for the sets $\pmb{X}(\pmb{x}'|\pmb{x})$. Such transition matrices can be thought of as direct product of simple matrices. That is they are resulted by concatenating matrices for single objects, $\pmb{X}_N=X_1..X_N$. Here $N$ is a size index which shows how many simple matrices are built into $\pmb{X}$. We drop such size index, and the size of a matrix $\pmb{X}$ can be inferred from the context. Sometimes such matrices include some extra numerical factor such as $\pmb{R}(\pmb{r}|\epsilon)=p_m\prod_{r\in\pmb{r}}R(r|\epsilon)$. This can be the case only about collective representation matrix $\pmb{R}$. We will make it specific if this happens. We use angle brackets $\langle.\rangle$ for an average over environmental states.\\
	We will normally drop convolution sign. Writing two matrices without space in between like $\pmb{X}\pmb{Y}(\pmb{x}|\pmb{y})$ or $XY(x|y)$  means convolution of those matrices $\pmb{X}\pmb{Y}(\pmb{x}|\pmb{y})=\sum_{\pmb{z}}\pmb{X}(\pmb{x}|\pmb{z})\pmb{Y}(\pmb{z}|\pmb{y})$. We also sometimes drop arguments altogether and simply write $\pmb{X}\pmb{Y}$ for convolution. Some times, for clarity of exposition we override this convention and make convolution sign specific. However, if there are arguments in between as in $X(x'|x)Y(y'|y)$ no convolution is involved. Also, when an individual and a collective matrix appear next to each other, as in $R\pmb{R}$, obviously no convolution is involved between $R$ and $\pmb{R}$, and this is simply a direct product: $R\pmb{R}=R(r|\epsilon)\pmb{R}(\pmb{r}|\epsilon)$. We normally specify this when it appears.\\
	We use $R$, $G$, and $C$, respectively for representation, production and comprehension matrices. $r$ and $\sigma$ are used for a representation and a signal. A lower index $t$ refers to time. An upper index $\rho$  is sometimes used for representations as in $r^{\rho}$. It refers to receiver's representation resulted from receiving and transforming a signal to a representation. Generally we use index $l$ to indicate local communication and index $g$ to indicate global communication. For example if a profile of signals is reached through local communication we add a lower index $l$ and write $\pmb{\sigma}_l$. $N$ is used for the population size. $N_t$ means population size in time $t$. $m$ is generally used for the number of signals in a profile. We use a dot $.$ for an empty set.\\
	We denote a CSS with representation matrix $R$, production matrix $G$, and comprehension matrix $C$, with $(R,G,C)$. The growth rate of a population equipped with a CSS $(R,G,C)$ is denoted by $\Lambda_{(R,G,C)}$ and the optimal growth by $\Lambda^*_{(R,G,C)}$. The value of a CSS is denoted by $V_{(R,G,C)}$. We show entropy by $H$ and mutual information by $I$. The joint probability densities over which the entropy is calculated is specified as lower index. For example $H_{p_{\epsilon}}(\epsilon)$ is the entropy of random variable $\epsilon$ with distribution $p_{\epsilon}$, and $H_{\pmb{CGR}p_{\epsilon}}(\epsilon|\pmb{r}^{\rho})$ is the conditional entropy of the environment given a profile of internal representations $\pmb{r}^{\rho}$, which is drawn according to the conditional distribution $\pmb{CGR}(\pmb{r}^{\rho}|\epsilon)$. In this case the joint probability density is $\pmb{CGR}p_{\epsilon}$, which is specified as a lower index on $H$.
	\section{Definition of Lyapunov exponent}
	\label{lyapunov}
	In our model we have two sources of stochasticity:  1. Stochasticities resulted from environmental stochasticity which results from the fact that the effective environment $\bar{\epsilon}_t$, is a stochastic process. In a model of global communication, the effective environment is composed of the environmental process $\epsilon_t$, and the profile of signals which is transmitted globally to a non-negligible fraction of the population. And, in a model with local communication, the effective environment is the same as the environmental process. 2. Stochasticity resulted from stochasticities present in linguistic interactions and individual decision making. These functions are done probabilistically and result in fluctuations in population size even if we fix effective environmental realization.
	One approach to deal with these two sources of stochasticity, is to define the Lyapunov exponent as:
	\begin{align}
	\bar{\Lambda}=\lim_{T \to \infty}\frac{1}{T}\log(\bar{N}(T)/N(0)).
	\label{def1}
	\end{align}
	Where $\bar{N}$ means an average over population stochasticities fixing the effective environmental sequence [\Rivoire]. 
	For a stationary environment, It can be shown that under rather general conditions (positivity of the Lyapunov exponent which is equivalent to the assumption of non-extinction, and a technical condition of stability which is trivially satisfied in our model) this Lyapunov exponent coincides with long term growth rate as we have defined in the main text [\Rivoire],
	\begin{align}
	\Lambda=\lim_{T \to \infty}\frac{1}{T}\log(N(T)/N(0)).
	\label{def2}
	\end{align}
	Where $N$ is the population size. Thus in the main text, and in the following, we have generally replaced expressions for the population size in the long term growth rate with expressions for mean population size. Intuitively this is because when the Lyapunov exponent is positive, the population grows exponentially and we are in the limit of large population size in which estimating population size with its mean becomes an exact treatment.\\
	\section{Local communication}
	\label{local}
	\subsection{Growth rate (derivation of equation (\hetmain))}
	\label{localgrowth}
	In this section, as in the main text, we consider a model with local communication. That is a model in which each individual who observe the world transmits a signal to its local neighbourhood. In the limit of large population size, such a local neighbourhood composes a negligible fraction of the community.\\
	For generality, we work with a rather more general growth model than introduced in the main text by taking the multiplication rates non-diagonal: $w_{\epsilon,s}\neq 0$ if $s\neq s_{\epsilon}$. Here $s_{\epsilon}$ is the best strategy in environment $\epsilon$. That is the strategy with highest multiplication rate, in environment $\epsilon$. In the diagonal model, all the strategies but the best one have multiplication rate $0$. At the end of calculations, we set the multiplication rate diagonal.\\
	We assume that in each environmental state, each individual observes the world state with probability $q$. In addition, each individual receives a profile of signals $\pmb{\sigma}_l$, in each environmental state $\epsilon$, according to a probability distribution $\pmb{N}_{l}(\pmb{\sigma}_l|\epsilon)$. The exact form of this probability distribution depends on the micro-structure of communication in the community and does not affect the following results. In section (\ref{miccomloc}) of this appendix, two examples of such a distribution for two specific models of local communication is given.
	The population evolves according to the equation:
	\begin{align}
	N_{t+1}=& \sum_{\pmb{r}_l^{\rho},s} w_{\epsilon_{t},s}(\sum_{  r}qA_1(s|r,\pmb{r}_l^{\rho})R(r|\epsilon_t)\nonumber\\&+(1-q)A_2(s|\pmb{r}^{\rho}_l))\pmb{C}_l\pmb{N}_l(\pmb{r}^{\rho}_l|\epsilon_t)N_t.
	\label{het1}
	\end{align}
	Where as usual the convolution sign is dropped: $\pmb{C}_l\pmb{N}_l(\pmb{r}_l^{\rho}|\epsilon_t)=\sum_{\pmb{\sigma}_l}\pmb{C}_l(\pmb{r}^{\rho}_l|\pmb{\sigma}_l)\pmb{N}_l(\pmb{\sigma}_l|\epsilon_t)$. Here $\sum_{\pmb{r}^{\rho}}$ is a summation over elements of the set and sets with different size $\sum_{\pmb{r}^{\rho}}=\sum_{r\in \pmb{r}^{\rho}}\sum_{\pmb{r}^{\rho}}'$. In writing this equation we have used the fact that when each individual observes the world with probability $q$, the mean and the variance of the number of individuals who observe the world both is proportional to $N_t$. Thus the ratio of mean to standard deviation is zero in the limit of large population size, and a fraction $q$ of the individuals observe the world directly. Hence, a ratio $qR(r|\epsilon)$ obtain representation $r$ through observation. In addition these individuals receive a profile of signals $\pmb{\sigma}_l$ according to $\pmb{N}_l(\pmb{\sigma}_l|\epsilon)$ and recover a profile of representations $\pmb
	{r}^{\rho}_l$ corresponding to it according to $\pmb{C}_l(\pmb{r}^{\rho}_l|\pmb{\sigma}_l)$. Consequently their internal state is $(r,\pmb{r}_l^{\rho})$ and choose strategy $s$ given their internal state according to their action matrix $A_1(s|r,\pmb{r}_l^{\rho})$. This corresponds to the first term $\sum_{r,\pmb{r}_l^{\rho}}qA(s|r,\pmb{r}^{\rho}_l)R(r|\epsilon_t)\pmb{C}_l\pmb{N}_l(\pmb{r}_l^{\rho}|\epsilon_t)N_t$ in Eq. (\ref{het1}). Others who compose a fraction $1-q$ of the population, do not make an observation and only receive a profile of signals $\pmb{\sigma}_l$ according to $\pmb{N}_l$. They transform it to a profile of representations according to $\pmb{C}_l$. Thus their internal state is $\pmb{r}_l^{\rho}$, and choose strategy $s$ with probability $A(s|\pmb{r}_l^{\rho})$. This corresponds to a fraction $\sum_{\pmb{r}^{\rho}_l}(1-q)A(s|\pmb{r}_l^{\rho}))\pmb{C}_l\pmb{N}_l(\pmb{r}^{\rho}_l|\epsilon_t)$ of the population and is the second term in Eq. (\ref{het1}). Therefore, altogether a fraction $\sum_{\pmb{r}_l^{\rho}} (\sum_{  r}qA_1(s|r,\pmb{r}^{\rho}_l)R(r|\epsilon_t)+(1-q)A_2(s|\pmb{r}_l^{\rho}))\pmb{C}_l\pmb{N}_l(\pmb{r}_l^{\rho}|\epsilon_t)$ of the population choose strategy $s$. These contribute to growth an amount $\sum_{\pmb{r}^{\rho}_l} w_{\epsilon_{t},s}(\sum_{  r}qA_1(s|r,\pmb{r}^{\rho}_l)R(r|\epsilon_t)+(1-q)A_2(s|\pmb{r}_l^{\rho}))\pmb{C}_l\pmb{N}_l(\pmb{r}^{\rho}_l|\epsilon_t)N_t$. Summing over all strategies, we have Eq. (\ref{het1}). \\
	In order to make the following expressions look less ugly, we introduce an abbreviated notation. We define:
	\begin{align}
	\pmb{N}_q(z,\pmb{r}_l^{\rho}|\epsilon,\pmb{\sigma}_l)=qR(z|\epsilon)\pmb{C}_l(\pmb{r}_l^{\rho}|\pmb{\sigma}_l)+(1-q)\delta_{z,.}\pmb{C}_l(\pmb{r}_l^{\rho}|\pmb{\sigma}_l).
	\label{het2}
	\end{align}
	Where $z$ is a random variable which can be either $r$ or an empty set $.$, and $R(z=r|\epsilon)=R(r|\epsilon)$ and $R(z=.|\epsilon)=0$. Here, we have introduced a delta function $\delta_{z,.}$ which is $1$ if $z$ is an empty set and is zero otherwise. The action matrix becomes $A(s|z,\pmb{r}_l^{\rho})$, which is equivalent to $A_1(s|r,\pmb{r}_l^{\rho})$ if $z=r$ and equivalent to $A_2(s|\pmb{r}_l^{\rho})$ if $z=.$. With this notation Eq. (\ref{het1}) can equivalently be written as:
	\begin{align}
	N_{t+1}=\sum_sw_{\epsilon_t,s}A\pmb{N}_q\pmb{N}_l(s|\epsilon_t)N_t.
	\label{het3}
	\end{align}
	Where, as usual, we have dropped convolution sign $A\pmb{N}_q\pmb{N}_l(s|\epsilon_t)=A*\pmb{N}_q*\pmb{N}_l(s|\epsilon_t)=\sum_{\pmb{r}_l^{\rho},\pmb{\sigma}_l}(\sum_rqA(s_{\epsilon_t}|r,\pmb{r}_l^{\rho})R(r|\epsilon_t)+(1-q)A(s_{\epsilon_t}|\pmb{r}_l^{\rho}))\pmb{C}_l(\pmb{r}_l^{\rho}|\pmb{\sigma}_l)\pmb{N}_l(\pmb{\sigma}_l|\epsilon_t)$. Iterating Eq. (\ref{het3}) from time $0$ to time $T-1$, we have for the population evolution equation:
	\begin{align}
	N_{T}=\prod_{t=0}^{T-1}  \sum_sw_{\epsilon_{t},s}A\pmb{N}_q\pmb{N}_l(s|\epsilon)N_0.
	\label{het4}
	\end{align}
	Dividing this by $N_0$, taking the logarithm, dividing by $T$, and then taking the large time limit we have for the long term growth rate:
	\begin{align}
	&\Lambda_{(A,G,C)}^l=\lim_{t \to \infty} \frac{1}{T}\sum_{t=0}^{T-1}\log(\sum_sw_{\epsilon_t,s}A\pmb{N}_q\pmb{N}_l(s|\epsilon_t)).
	\label{het5}
	\end{align}
	Using the stationarity of the process to replace time summation with a summation over states we have:
	\begin{align}
	&\Lambda_{(A,G,C)}^l=\sum_{\epsilon}p_{\epsilon}\log(\sum_sw_{\epsilon,s}A\pmb{N}_q\pmb{N}_l(s|\epsilon)).
	\label{het6}
	\end{align}
	Taking the multiplication rate diagonal, we arrive at:
	\begin{align}
	&\Lambda_{(A,G,C)}^l=\sum_{\epsilon}p_{\epsilon}\log(w_{\epsilon}A\pmb{N}_q\pmb{N}_l(s_{\epsilon}|\epsilon)).
	\label{het06}
	\end{align}
	Which is Eq. (\hetmain) in the main text.
	\subsection{Optimal growth (derivation of equation (\decay))}
	\label{localoptimal}
	The optimal growth rate is achieved by optimizing Eq. (\ref{het06}) with respect to action matrix subject to the constraint $\sum_{\epsilon}A(s_{\epsilon}|z,\pmb{r}^{\rho})=1$. Formally, we can write the resulting optimal growth rate in the following form:
	\begin{align}
	&\Lambda_{(R,G,C)}^{l*}=\langle\log w_{\epsilon}\rangle-H^{\pmb{N}_q\pmb{N}_l}_{p_{\epsilon}}(\epsilon|z,\pmb{r}_l^{\rho}).
	\label{het7}
	\end{align}
	where
	\begin{align}
	H^{\pmb{N}_q\pmb{N}_l}_{p_{\epsilon}}(\epsilon|z,\pmb{r}^{\rho}_l)=\min_{A} \sum_{\epsilon}-p_{\epsilon}\log(A\pmb{N}_q\pmb{N}_l(s_{\epsilon}|\epsilon)).
	\label{het8}
	\end{align}
	\\By subtracting Eq. (\ref{het7}) from the growth rate of the same population without a CSS in equation (\lambd) in the main text, for the value of CSS we derive:
	\begin{align}
	V^{l}_{(R,G,C)}=\Lambda_{(R,G,C)}^{l*}-\Lambda_0^*=I^{\pmb{N}_q\pmb{N_l}}_{p_{\epsilon}}(\epsilon;z,\pmb{r}_l^{\rho}).
	\label{het9}
	\end{align}
	Where we have defined:
	\begin{align}
	I^{\pmb{N}_q\pmb{N}_l}_{p_{\epsilon}}(\epsilon;z,\pmb{r}^{\rho}_l)=H_{p_{\epsilon}}(\epsilon)-H^{\pmb{N}_q\pmb{N}_l}_{p_{\epsilon}}(\epsilon|z,\pmb{r}_l^{\rho})
	\end{align}
	\subsection{Beneficial effect of heterogeneity (derivation of equation (\ineqhet))}
	\label{localhet2}
	Here, we want to derive equation (\ineqhet) in the main text. In addition, we will give an interpretation to it, based on the way in which various sources of heterogeneity enhance the growth.\\
	Putting $\pmb{N}_q$ from Eq. (\ref{het2}) in Eq. (\ref{het6}), we can write the growth rate in the presence of local communication, Eq. (\ref{het6}), in a more explicit form:
	\begin{align}
	& \Lambda_{(A,G,C)}^l=\sum_{\epsilon}p_{\epsilon}\log(\sum_{s,r,\pmb{r}_l^{\rho}} w_{\epsilon,s}(qA_1(s|r,\pmb{r}_l^{\rho})R(r|\epsilon)\nonumber\\&+(1-q)A_2(s|\pmb{r}_l^{\rho}))\pmb{C}_l\pmb{N}_l(\pmb{r}_l^{\rho}|\epsilon)).
	\label{ineq9}
	\end{align}
	Using Jensen's inequality to drag $q\pmb{N}_l$ and $(1-q)\pmb{N}_l$ outside of the argument of the logarithm, we have:
	\begin{align}
	& \Lambda_{(A,G,C)}^l \geq q\sum_{\pmb{\sigma}_l,\epsilon}\pmb{N}_l(\pmb{\sigma}_l^{\rho}|\epsilon)p_{\epsilon}\log(\sum_{s,r,\pmb{r}_l^{\rho}}w_{\epsilon,s} A_1(s|r,\pmb{r}_l^{\rho})R(r|\epsilon)\nonumber\\&\pmb{C}_l(\pmb{r}_l^{\rho}|\pmb{\sigma}^{\rho}_l))+(1-q)\sum_{\pmb{\sigma},\epsilon}\pmb{N}_l(\pmb{\sigma}_l^{\rho}|\epsilon)p_{\epsilon}\log(\sum_{s,\pmb{r}_l^{\rho}}w_{\epsilon,s} A_2(s|\pmb{r}_l^{\rho})\nonumber\\&\pmb{C}_l(\pmb{r}_l^{\rho}|\pmb{\sigma}_l))=\Lambda'^{l}_{(A,G,C)}
	\label{ineq10}
	\end{align}
	$\Lambda'^l_{(A,G,C)}$ is the growth rate of a population in which the whole population receives a signal $\pmb{\sigma}_l$ according to $\pmb{N}_l$ globally, in addition, in each time step, the whole population observe the world state with probability $q$, and do not make an observation with probability $1-q$.\\
	Now we use the fact that $\sum_{r,\pmb{r}^{\rho}_l} R(r|\epsilon)\pmb{C}_l(\pmb{r}_l^{\rho}|\pmb{\sigma}_l)=1$ and $\sum_{\pmb{r}^{\rho}_l} \pmb{C}_l(\pmb{r}_l^{\rho}|\pmb{\sigma}_l)=1$, to use Jensen's inequality to take these terms outside of the argument of logarithms in Eq. (\ref{ineq10}). Doing so we have:
	\begin{align}
	& \Lambda'^l_{(A,G,C)} \geq q\sum_{\epsilon,r,\pmb{r}_l^{\rho}}R(r|\epsilon)\pmb{C}_l\pmb{N}_l(\pmb{r}_l^{\rho}|\epsilon)p_{\epsilon}\log(\sum_sw_{\epsilon,s} \nonumber\\&A_1(s|r,\pmb{r}^{\rho}_l))+(1-q)\sum_{\epsilon,\pmb{r}_l^{\rho}}\pmb{C}_l\pmb{N}_l(\pmb{r}_l^{\rho}|\epsilon)p_{\epsilon}\log(\sum_sw_{\epsilon,s}\nonumber\\& A_2(s|\pmb{r}_l^{\rho}))=\Lambda''^l_{(A,G,C)}.
	\label{ineq11}
	\end{align}
	$\Lambda''^l_{(A,G,C)}$ is the growth rate of a population in which the whole population with probability $q$ receives an internal state $(r,\pmb{r}_l^{\rho})$ by making an observation through a channel $R(r|\epsilon)\pmb{C}_l\pmb{N}_l(\pmb{r}_l^{\rho}|\epsilon)$, and with probability $1-q$ receives an internal state $\pmb{r}_l^{\rho}$ by observation through a channel $\pmb{C}_l\pmb{N}_l(\pmb{r}_l^{\rho}|\epsilon)$, homogeneously.
	From Eq. (\ref{ineq10}) and Eq. (\ref{ineq11}) we have:
	\begin{align}
	\Lambda^l_{(A,G,C)} \geq \Lambda''^l_{(A,G,C)}.
	\label{ineq12}
	\end{align}
	We note that $\Lambda''^l_{(A,G,C)}$ is the growth rate of a hypothetical population who makes observations from the world using the same channels used in a CSS with local communication. As several steps of the derivations show, various sources of heterogeneity such as in observation through the channel $R$, in production of signals, comprehension of signals, and heterogeneity resulted from the population being composed of those who observe and who does not, cause the value of CSS be larger than this hypothetical population. All these sources of heterogeneity can be thought of as sources which amplify the value of acquiring information through a CSS. We note that we have derived this inequality in a general situation. That is for non-diagonal multiplication rate and all the strategies not only the optimal one.\\
	Equation (\ineqhet) in the main text is derived for optimal growth of the diagonal model. By optimizing Eq. (\ref{ineq12}) we have:
	\begin{align}
	\Lambda^{l*}_{(A,G,C)} \geq \Lambda''^{l*}_{(A,G,C)}.
	\label{ineq14}
	\end{align}
	We can derive the growth rate of the diagonal model by setting $w_{\epsilon,s}=0$ if $s\neq s_{\epsilon}$ in Eq. (\ref{ineq11}):
	\begin{align}
	&\Lambda''^l_{(A,G,C)}=q\sum_{\epsilon,r,\pmb{r}_l^{\rho}}R(r|\epsilon)\pmb{C}_l\pmb{N}_l(\pmb{r}_l^{\rho}|\epsilon)p_{\epsilon}\log(w_{\epsilon} A_1(s_{\epsilon}|r,\pmb{r}^{\rho}_l))\nonumber\\&+(1-q)\sum_{\epsilon,\pmb{r}_l^{\rho}}\pmb{C}_l\pmb{N}_l(\pmb{r}_l^{\rho}|\epsilon)p_{\epsilon}\log(w_{\epsilon} A_2(s_{\epsilon}|\pmb{r}_l^{\rho})).
	\label{ineq014}
	\end{align}
	Now optimizing $\Lambda''^l_{(A,G,C)}$ subject to the constraints $\sum_{s}A_1=1$ and $\sum_{s}A_2=1$, we find
	\begin{align}
	\Lambda''^{l*}_{(A,G,C)}=&q(\langle \log w_{\epsilon}\rangle - H_{R(r|\epsilon)\pmb{C}_l\pmb{N}_l(\pmb{r}_l^{\rho}|\epsilon)}(\epsilon|r,\pmb{r}_l^{\rho}))\nonumber\\&+(1-q)(\langle\log w_{\epsilon}\rangle - H_{\pmb{C}_l\pmb{N}_l(\pmb{r}_l^{\rho}|\epsilon)}(\epsilon|\pmb{r}_l^{\rho})).
	\label{ineq13}
	\end{align}
	From Eq. (\ref{het7}), Eq. (\ref{ineq13}), and Eq. (\ref{ineq14}) we have:
	\begin{align}
	H^{\pmb{N}_q\pmb{N}_l}_{p_{\epsilon}}(\epsilon|z,\pmb{r}_l^{\rho}) \leq &q H_{R(r|\epsilon)\pmb{C}_l\pmb{N}_l(\pmb{r}^{\rho}|\epsilon)}(\epsilon|r,\pmb{r}^{\rho}_l)\nonumber\\&+(1-q) H_{\pmb{C}_l\pmb{N}_l(\pmb{r}_l^{\rho}|\epsilon)}(\epsilon|\pmb{r}_l^{\rho}).
	\label{ineq15}
	\end{align}
	Which is the entropic inequality mentioned in equation (\ineqhet) in the main text.
	By subtracting the growth rate of the base line model in equation (\lambd) in the main text, from Eq. (\ref{ineq14}) and using Eq. (\ref{ineq15}), we have the following inequality for the value of CSS with local communication $V^l_{(A,G,C)}$:
	\begin{align}
	V^l_{(A,G,C)} \geq &q I_{R(r|\epsilon)\pmb{C}_l\pmb{N}_l(\pmb{r}_l^{\rho}|\epsilon)}(\epsilon;r,\pmb{r}_l^{\rho})\nonumber\\&+(1-q) I_{\pmb{C}_l\pmb{N}_l(\pmb{r}_l^{\rho}|\epsilon)}(\epsilon;\pmb{r}_l^{\rho}).
	\label{ineq16}
	\end{align}
	Which is the inequality for extended mutual information in equation (\ineqhet) in the main text.\\
	We note that as the extent of communications increase (i.e. the average number of signals received by an individual increases), the conditional entropy of the environment given the internal states decreases. Consequently the write hand side of Eq. (\ref{ineq15}) decreases. As $H^{\pmb{N}_q\pmb{N}_l}_{p_{\epsilon}}(\epsilon|z,\pmb{r}_l^{\rho})$ is bounded by this conditional entropy, this also decreases and tends to zero for high extent of communications. In the same way, as by increasing the extent of communications the profile of received internal representations determines the environment with high accuracy and removes all the uncertainty, the mutual informations in the right hand side of Eq. (\ref{ineq16}) tend to entropy of the environment. As $V^l_{(A,G,C)}$ is bounded between entropy of the environment and the right hand side of Eq. (\ref{ineq16}), it thus tends to environmental entropy. Intuitively this phenomenon results from the fact that by increasing the extent of communication, as the number of signals contained in a profile of representations increases, it specify environmental state with better and better accuracy.
	\subsection{Beneficial effect of heterogeneity (derivation of equation (\hemain))}
	\label{localhet1}
	To show the beneficial effect of heterogeneity, we consider the same model, but instead of receiving a profile of signals $\pmb{\sigma}_l$ according to a probability $\pmb{N}_l(\pmb{\sigma}_l|\epsilon)$, through local communications, individuals receive a profile of signals $\pmb{\sigma}$ according to the same distribution $\pmb{N}_l$ but by global communication. That is in each environmental state, a profile of signals is drawn according to the same distribution and broadcasted to the whole population by some individuals. Here again to keep calculations general, we work with a non-diagonal model and set the multiplication rate diagonal at the end of the calculations.\\
	To be more specific, we consider a population in which each individual observes the world with probability $q$. In addition the whole population receive a profile of signals $\pmb{\sigma}$ homogeneously, according to $\pmb{N}_l(\pmb{\sigma}|\epsilon)$, transmitted by a subset of individuals who have observed the world state. Thus, a fraction q reach the internal state $(r,\pmb{r}^{\rho})$, with probability $R(r|\epsilon)\sum_{\pmb{\sigma}}\pmb{C}(\pmb{r}^{\rho}|\pmb{\sigma})\pmb{N}_l(\pmb{\sigma}|\epsilon)$. Consequently, they choose their strategy according to $A_1(s|r,\pmb{r}^{\rho})$. The rest, a fraction $1-q$ of the population, do not observe the world and form an internal state based solely on the signals they receive. Their internal state will be $\pmb{r}^{\rho}$ with probability $\sum_{\pmb{\sigma}}\pmb{C}(\pmb{r}^{\rho}|\pmb{\sigma})\pmb{N}_l(\pmb{\sigma}|\epsilon)$. Consequently they choose strategy $s$ with probability $A_2(s|\pmb{r}^{\rho})$. Given all these the population evolution equation will be:
	\begin{align}
	N_{t+1}=& \sum_{s,\pmb{r}^{\rho}} w_{\epsilon_{t},s}(\sum_rqA(s|r,\pmb{r}^{\rho})R(r|\epsilon_t)\nonumber\\&+(1-q)A(s|\pmb{r}^{\rho}))\pmb{C}(\pmb{r}^{\rho}|\pmb{\sigma}_t)N_t.
	\label{he1}
	\end{align}
	Using the abbreviated notation introduced before in this section, this can be written:
	\begin{align}
	N_{t+1}=\sum_sw_{\epsilon_t,s}A\pmb{N}_q(s|\epsilon_t,\pmb{\sigma}_t)N_t.
	\label{he2}
	\end{align}
	Where, as always we have dropped the convolution sign $A\pmb{N}_q(s|\pmb{\sigma}_t)=A*\pmb{N}_q(s|\pmb{\sigma}_t)$.
	Iterating this we have for the population size in time $T$
	\begin{align}
	N_{T}=\prod_{t=0}^{T-1} \sum_s w_{\epsilon_{t},s}A\pmb{N}_q(s|\epsilon_t,\pmb{\sigma}_t)N_0.
	\label{he3}
	\end{align}
	By dividing by $N_0$, taking the logarithm, dividing by $T$, and taking the large time limit we have for the growth rate:
	\begin{align}
	\Lambda^{'}=\lim_{T\to\infty}\sum_{t=0}^{T-1}\frac{1}{T}\log(w_{\epsilon}A\pmb{N}_q(s_{\epsilon_t}|\epsilon_t,\pmb{\sigma}_t).
	\label{he4}
	\end{align}
	By using the stationarity of the environmental process we have for the growth rate:
	\begin{align}
	\Lambda^{'}=\sum_{\epsilon,\pmb{\sigma}}\pmb{N}_l(\pmb{\sigma}|\epsilon)p_{\epsilon}\log(\sum_sw_{\epsilon,s}A\pmb{N}_q(s|\epsilon,\pmb{\sigma})).
	\label{he5}
	\end{align}
	By setting the multiplication rate diagonal, this becomes equation (\hemain) in the main text. By using Jensen's inequality to take $\sum_{\pmb{\sigma}}\pmb{N}_l$ inside of the argument of logarithm, we can easily see that:
	\begin{align}
	\Lambda_{(R,G,C)}^l\geq\Lambda'.
	\label{he005}
	\end{align}
	This shows the positive effect of heterogeneity. Because, drawing a profile of signals according to the distribution $\pmb{N}_l$ in each environmental state and transmitting it to the whole population homogeneously leads to lower growth than when each individual receives a profile $\pmb{\sigma}_l$, according to the same distribution $\pmb{N}_l$ heterogeneously. Thus, in the second model individuals receive a profile with the same information content, but heterogeneously, such that each individual can receive a different profile. We see that this heterogeneity enhances the growth.\\
	Even though we have not used the optimal growth of this model in the main text, we calculate the optimal growth of this model as well for completeness. First by setting the multiplication rate diagonal from Eq. (\ref{he5}), we get for the growth rate of this model in the diagonal case:
	\begin{align}
	\Lambda^{'}=\sum_{\epsilon,\pmb{\sigma}}\pmb{N}_l(\pmb{\sigma}|\epsilon)p_{\epsilon}\log(w_{\epsilon}A\pmb{N}_q(s_{\epsilon}|\epsilon,\pmb{\sigma})).
	\label{he05}
	\end{align}
	Optimizing Eq. (\ref{he05}) with respect to $A(s|z,\pmb{r}^{\rho})$ subject to the constraint $\sum_{s}A(s|z,\pmb{r}^{\rho})=1$, we have for the optimal growth rate:
	\begin{align}
	&\Lambda^{'*}=\langle\log w_{\epsilon}\rangle-H^{\pmb{N}_q}_{\pmb{N}_{l}p_{\epsilon}}(\epsilon|r,\pmb{r^{\rho}}).
	\label{he6}
	\end{align}
	Where $H^{\pmb{N}_q}_{\pmb{N}_{l}p_{\epsilon}}(\epsilon|r,\pmb{r}^{\rho})=\min_{A} \sum_{\epsilon,\pmb{\sigma}}-\pmb{N}_l(\pmb{\sigma}|\epsilon)p_{\epsilon}\log(A\pmb{N}_q(s_{\epsilon}|\\\epsilon,\pmb{\sigma}))$.
	And the value of CSS in this case is derived by subtracting the optimal growth of a population without CSS in equation (\lambd) in the main text from Eq. (\ref{he6}):
	\begin{align}
	V^{'*}=I^{\pmb{N}_q}_{\pmb{N}_{l}p_{\epsilon}}(\epsilon;r,\pmb{r}^{\rho})=H_{p_{\epsilon}}(\epsilon)-H^{\pmb{N}_q}_{\pmb{N}_{l}p_{\epsilon}}(\epsilon|r,\pmb{r}^{\rho}).
	\label{he7}
	\end{align}
	\begin{figure}
		\includegraphics[width=1\linewidth]{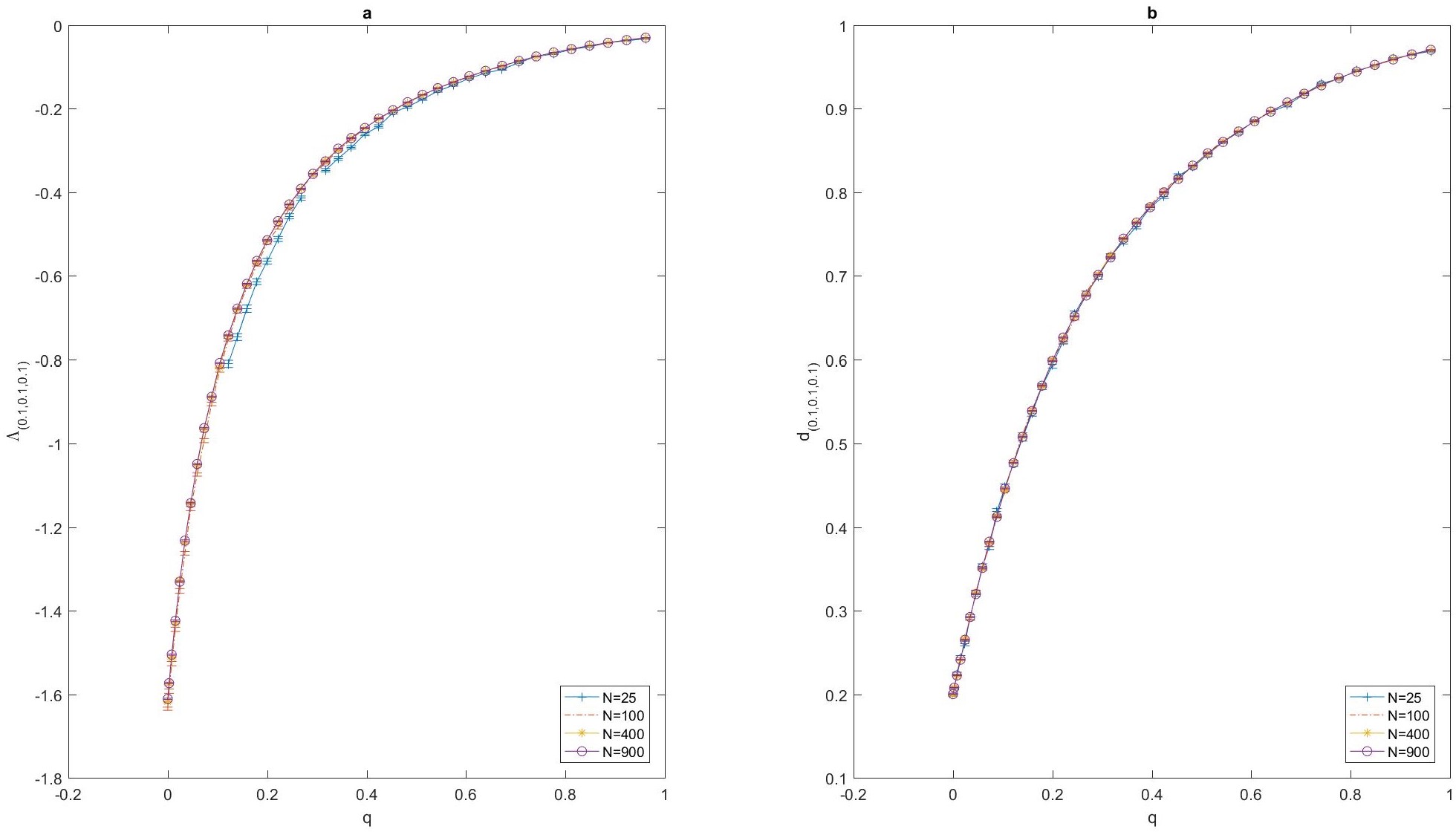}
		\caption{(a) Growth rate and, (b) the fraction of individuals who choose the correct strategy $d$, calculated on different network sizes, versus observation probability. We see that $d$, and the resulting growth rate are independent of the network size, provided that the network is large enough so extinction does not occur. Here for $N=25$, for many values of $q$ extinction occurs. However, for population sizes as large as $N\geq100$ the growth rates calculated based on different fixed network sizes are the same within error bars. The quantity calculated as growth rate here is given in Eq. (\ref{d}), setting $T=50$. Averages and standard errors are calculated based on a sample of $R=40$ simulations. An extinction in the graph corresponds to at least one extinction event in this sample. The presented values for $d$ and its standard errors are calculated based on a sample of $T=2000$ realizations. As with our parametrization of the CSS matrices all the states are symmetric, the result is independent of the environmental state.}
		\label{figNq}
	\end{figure}
	\begin{figure*}
	\includegraphics[width=1\linewidth]{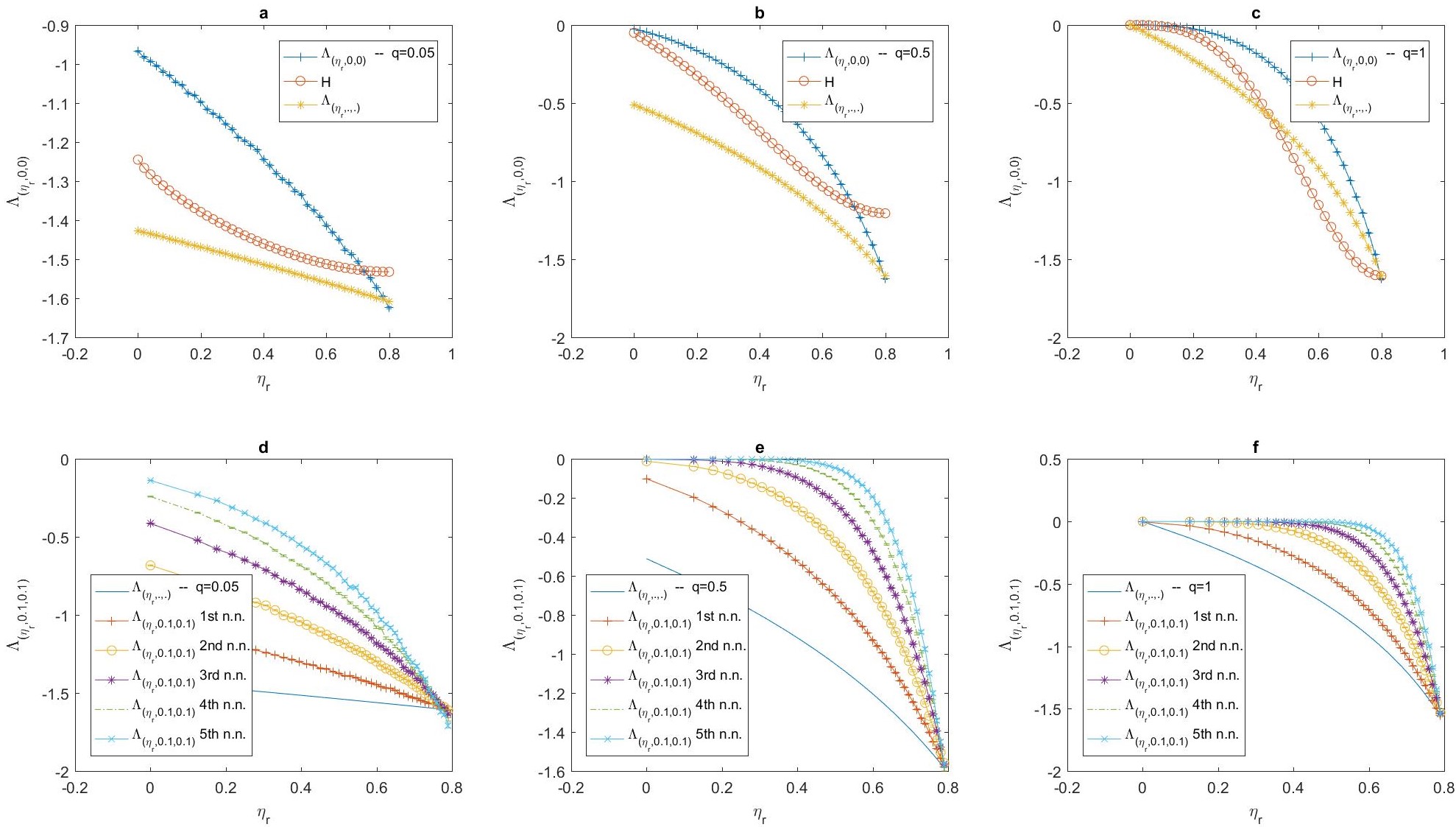}
	\caption{(a), (b), and (c): The growth rate on a square lattice with nearest neighbour interactions for a population using a majority rule as its decision making strategy, as a function of $\eta_r$, for different values of observation probabilities $q$, is plotted. The lower bound derived in the text, $H$, and the growth rate of a population with the same representation channel but without communication is plotted as well. We see that the bound derived for optimal growth rate, is not respected for high levels of noise in representation and $q\neq 1$, when individuals use a majority rule as their strategy. (d), (e), and (f): Growth rate on a square lattice for up to $l$th nearest neighbour interactions as a function of noise in representation for different $q$s is plotted. The growth rate approaches maximum growth exponentially as the noise in representation decreases near maximum noise level. The speed of increase in growth rate by noise reduction decreases however for low values of $q$. Besides, increasing the extent of communications increases growth such that the population can remove all the environmental uncertainty even for high levels of noise in representation, for moderate to large values of $q$.}
	\label{fig23Rq}
\end{figure*}
\begin{figure*}
	\includegraphics[width=1\linewidth]{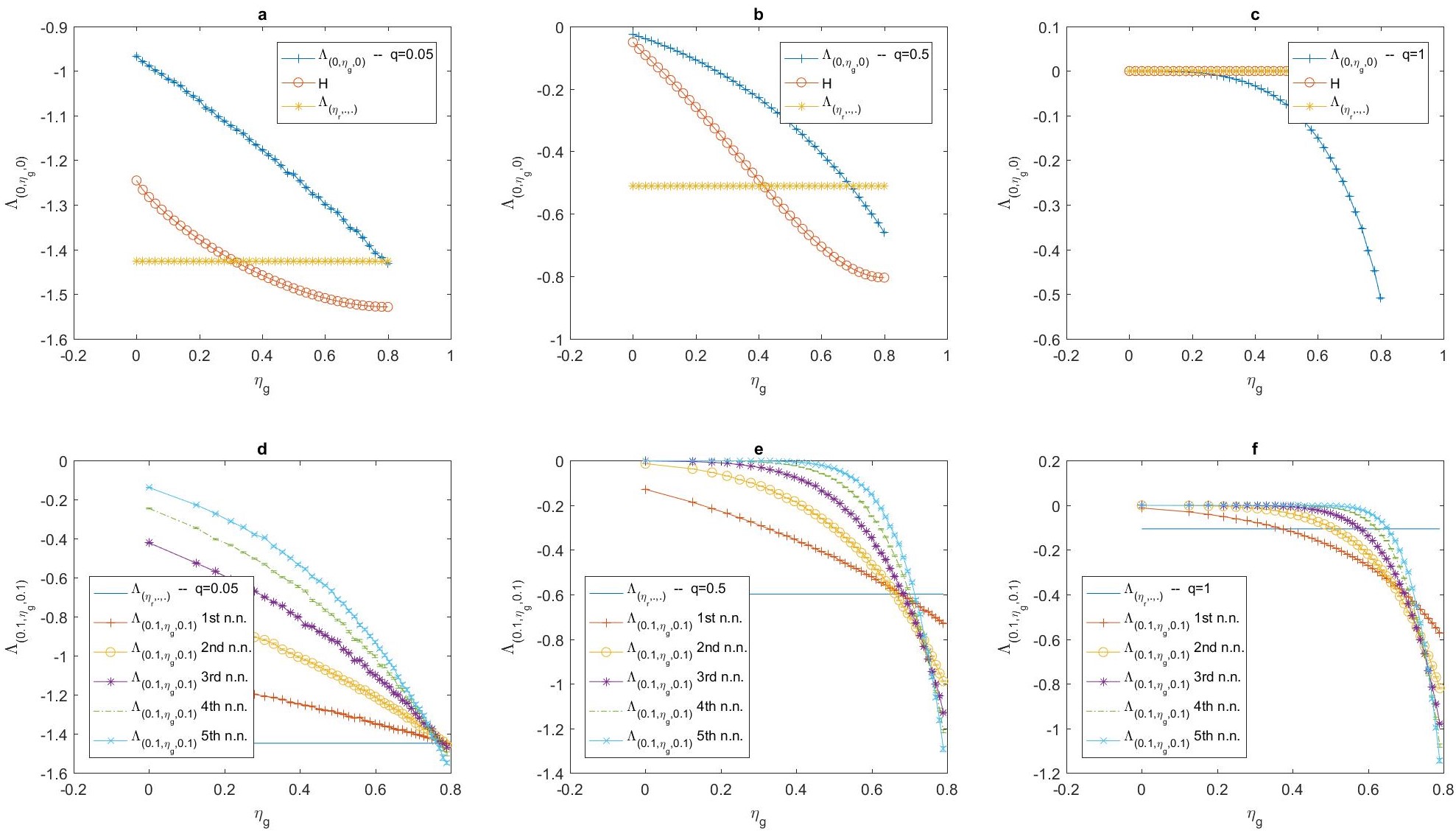}
	\caption{(a), (b), and (c): Growth rate on a square lattice with nearest neighbour interactions for a population using a majority rule as its decision making strategy, versus noise in production $\eta_g$, for different values of observation probability $q$. The lower bound derived in the text, $H$, and the growth rate of a population with the same representation channel but without communication is plotted as well. As can be seen in the plots, by increasing noise in communication, the value of communication decreases. (d), (e), and (f): Growth rate on a square lattice for up to $l$th nearest neighbour interactions as a function of noise in production, for different values of $q$, is plotted. The growth rate approaches maximum growth exponentially as the noise in production decreases near maximum noise level. Besides, increasing the extent of communications increases growth such that the population can remove all the environmental uncertainty even for high levels of noise in production, for moderate to large values of $q$.}
	\label{fig23Pq}
\end{figure*}
	\section{An example of the effective signal $O$}
	As an example of the effective signal $o$, consider a CSS given by $3$ environmental states, $2$ representation and signals, and the following conditional distribution:
	\begin{align}
	&CGR(r_1|\epsilon_1)=e_1\quad CGR(r_1|\epsilon_2)=CGR(r_1|\epsilon_3)=e_2\nonumber\\&
	CGR(r_2|\epsilon_1)=1-e_1 \quad CGR(r_2|\epsilon_2)=CGR(r_2|\epsilon_3)=1-e_2
	\end{align} 
	Where $e_1\neq e_2$. In this CSS, we can define the effective signal $o$ with two values $o_1$ and $o_2$. The signal $o_1$ can be thought of as a series of representations where a fraction $e_1$ of the representations are $r_1$ and a fraction $1-e_1$ are $r_2$, and the signal $o_2$ can be thought of as a series of representations where a fraction $e_2$ of the representations are $r_1$ and a fraction $1-e_2$ are $r_2$. This signal divides the environmental states into two classes: the class corresponding to $o_1$ is $\epsilon_1$, and the class corresponding to $o_2$ is $\{\epsilon_2,\epsilon_3\}$. We have for its conditional distributions $O(\epsilon_1|o_1)=1$ and $O(\epsilon_2|o_1)=0$, and $O(\epsilon_3|o_1)=0$. $O(\epsilon_1|o_2)=0$ and $O(\epsilon_2|o_2)=\frac{p_{\epsilon_2}}{p_{\epsilon_2}+p_{\epsilon_3}}$, and $O(\epsilon_3|o_2)=\frac{p_{\epsilon_3}}{p_{\epsilon_2}+p_{\epsilon_3}}$. We derive for the conditional entropy $H(\epsilon|o)=(p_{\epsilon_2}+p_{\epsilon_3})\log(p_{\epsilon_2}+p_{\epsilon_3})-p_{\epsilon_2}\log p_{\epsilon_2}-p_{\epsilon_3}\log p_{\epsilon_3}$. And for the mutual information between the effective signal $o$ and environment $\epsilon$, which is equal to the value of the CSS, we have $V_{(R,G,C)}=I(\epsilon;r)=(1-p_{\epsilon_1})\log(1-p_{\epsilon_1})-p_{\epsilon_1}\log p_{\epsilon_1}$.
	\section{Models for micro-structure of local communications}
	\label{miccomloc}
	We introduce two models for micro-structures of communication. The first one consider a case when the local communication results from spatial structure, and the second one considers a case when local communication results from a network of communications.
	\subsection{spatial structure}
	As a model for the detailed structure of local linguistic interactions we consider a case where each individual can detect the state of the world with probability $q$ and transmits a signal to others. The signal can propagate only as far as a characteristic length scale $l$. Thus, assuming the population lives in a $2$ dimensional landscape, only those individuals residing in a circle with radius $l$ centred around the transmitter can receive the signal.
	Assuming that the population occupies the landscape with density $\rho$, there live a total of $n_{l}=\pi l^2 \rho$ individuals in this circle. The probability that $n_{\pmb{\sigma}}$ individuals among this total $n_{l}$ individuals transmit a signal is given by the binomial distribution $P(n_{\pmb{\sigma}})=q^{n_{\pmb{\sigma}}} (1-q)^{n_{l}-n_{\pmb{\sigma}}}\binom{n_{l}}{n_{\pmb{\sigma}}}$. Each such profile is composed of a specific set of signals $\pmb{\sigma}$ with probability $\pmb{N}_{l}(\pmb{\sigma}|\epsilon)=q^{n_{\pmb{\sigma}}} (1-q)^{n_{l}-n_{\pmb{\sigma}}} \binom{n_{l}}{n_{\pmb{\sigma}}} 
	\sum_{r\in\pmb{r}}\prod_{r\in\pmb{r},\sigma\in\pmb{\sigma}}G(\sigma|r)$ $R(r|\epsilon)$, which in the interesting limit $q\to 0$ and $n_{l}\to \infty$ such that $qn_{l}=\lambda$ simplifies to $\pmb{N}_{l}(\pmb{\sigma}|\epsilon)=\exp(-\lambda)\frac{\lambda^{n_{\pmb{\sigma}} }}{n_{\pmb{\sigma}}!}
	\sum_{r\in\pmb{r}}\prod_{r\in\pmb{r},\sigma\in\pmb{\sigma}}G(\sigma|r)$ $R(r|\epsilon)$.\\
	\subsection{network structure}
	Assuming that the population resides on a communication network with degree distribution $p(k)$, we argue as follows. An individual with degree $k$, has $k$ neighbours. Each neighbour transmits a signal with probability $q$. Thus the probability that such individual receives a profile of signals with $n_{\pmb{\sigma}}$ signals, is: $q^{n_{\pmb{\sigma}}} (1-q)^{k-n_{\pmb{\sigma}}} \binom{k}{n_{\pmb{\sigma}}}$. Each such profile is composed of a specific set of signals with probability $q^{n_{\pmb{\sigma}}} (1-q)^{k-n_{\pmb{\sigma}}} \binom{k}{n_{\pmb{\sigma}}} 
	\prod_{\sigma\in\pmb{\sigma}}GR(\sigma|\epsilon)$. Averaging over the degree distribution we derive $\pmb{N}_{l}(\pmb{\sigma}|\epsilon)=\sum_k p(k)q^{n_{\pmb{\sigma}}} (1-q)^{k-n_{\pmb{\sigma}}} \binom{k}{n_{\pmb{\sigma}}} 
	\prod_{\sigma\in\pmb{\sigma}}GR(\sigma|\epsilon)$.\\
	We note that a necessary condition for the validity of the calculation given in the main text in the case that local communication results from network structure, is that for every $k$, we have $Np(k) \gg 1$. This is so as in this case statistical fluctuations around mean of values become negligible compared to mean. This holds for example on a lattice or a random network.
\section{Simulations for local communication}
\begin{figure*}
	\includegraphics[width=1\linewidth]{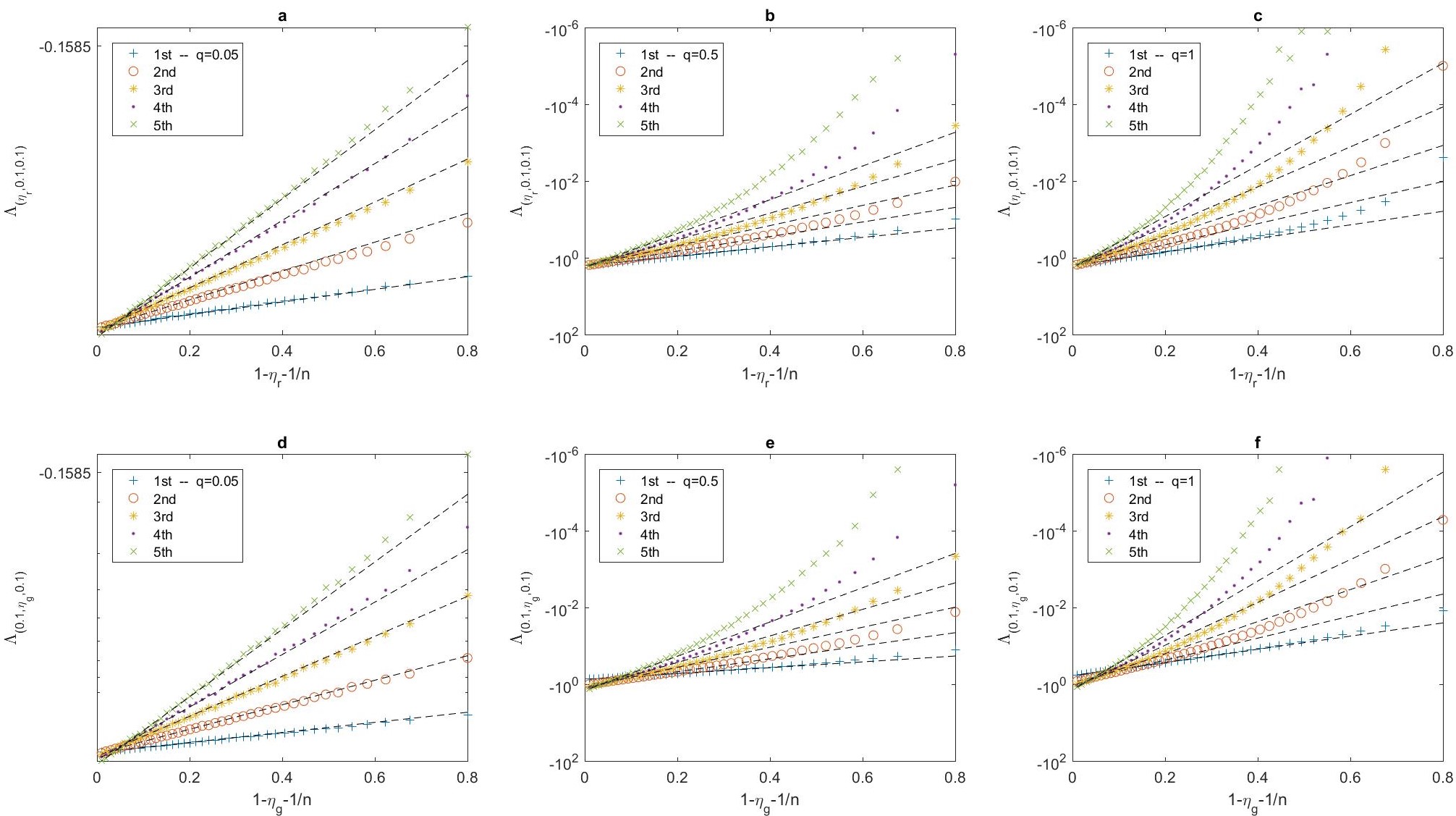}
	\caption{(a), (b), and (c): Log linear plot of the growth rate versus $1-\eta_r-\frac{1}{n}$, for different observation probabilities $q$. This is the probability by which a correct representation occurs above a uniformly random representation formation. We see that the growth rate increases exponentially as the probability of correct representation formation increases beyond a uniformly random representation formation matrix. The slope of the fit increases with $q$ and approaches to a $q$-independent value as $q$ increases. (d), (e), and (f): Log linear plot of the growth rate versus $1-\eta_r-\frac{1}{n}$ for different values of $q$. This is the probability by which a correct signal is produced above a uniformly random signal production. We see that the growth rate increases exponentially as the probability of correct signal production increases beyond a uniformly random signal production matrix. The slope of the fit increases with $q$ and approaches to a $q$-independent value as $q$ increases.}
	\label{figexpq}
\end{figure*}
\begin{figure*}
	\includegraphics[width=1\linewidth]{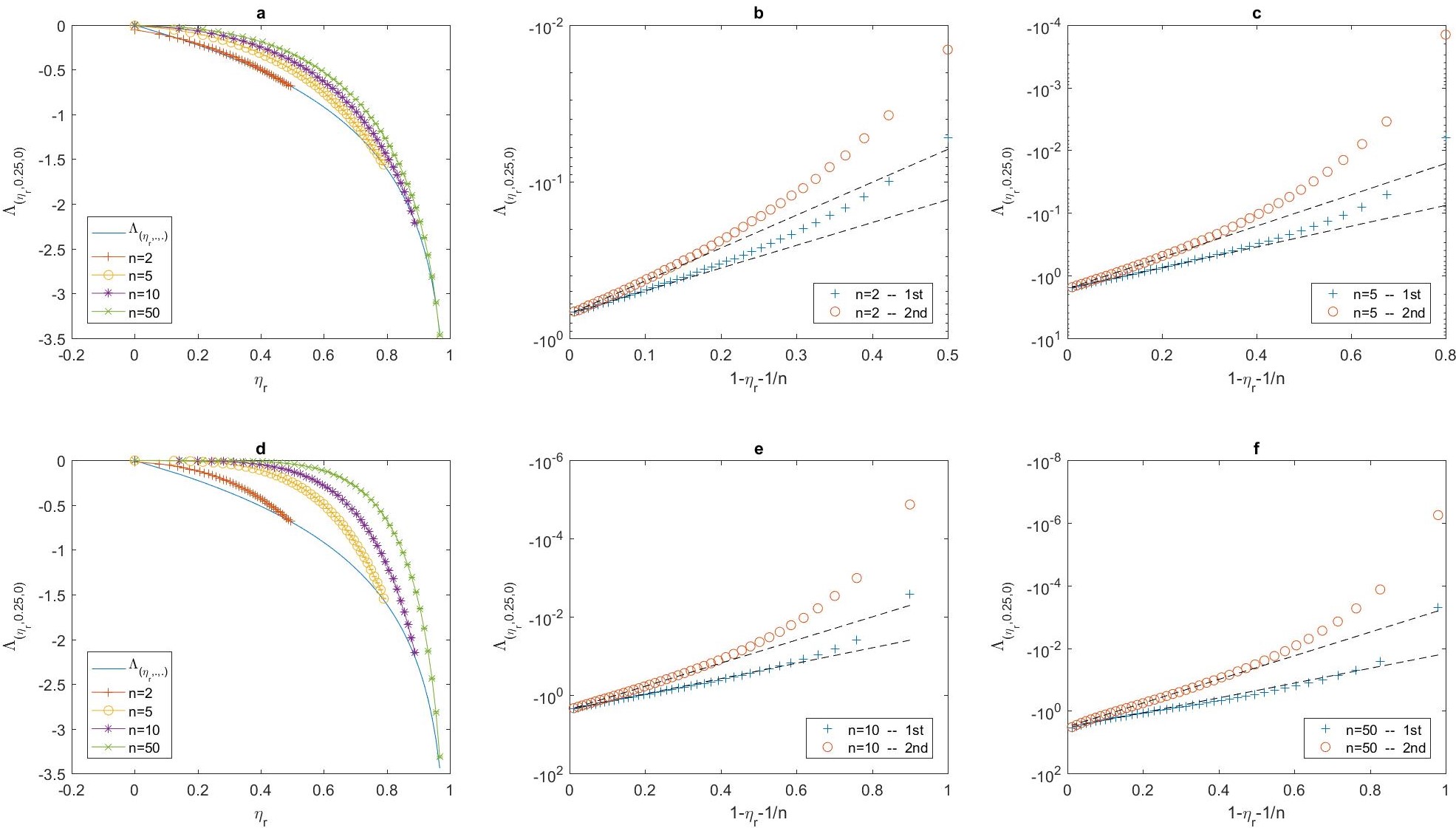}
	\caption{(a), (d): The growth rate on a square lattice with nearest neighbour interactions for a population using a majority rule as its decision making strategy, as a function of $\eta_r$, for different number of states $n$, is plotted. The growth rate of a population with the same representation channel but without communication is plotted as well. Here $q=1$. In (a) $l=1$ and in (d) $l=2$. We see that for the chosen noise level in communication , the value of communication is always positive and increases with $l$. As $n$ increases, the growth rate decreases for high noise levels. This is so because in high noise levels the number of individuals who choose the correct strategy approaches $\frac{1}{n}$ which decreases with $n$. (b), (c), (e), and (f): Log linear plot of the growth rate versus $1-\eta_r-\frac{1}{n}$ for different values of $n$. This is the probability by which a correct representation is formed above a uniformly random signal production. We see that the growth rate increases exponentially as the probability of correct signal production increases beyond a uniformly random signal production matrix for all $n$s. The slope of the fit is generally different for different values of $n$.}
	\label{figncRep}
\end{figure*}
\begin{figure*}
	\includegraphics[width=1\linewidth]{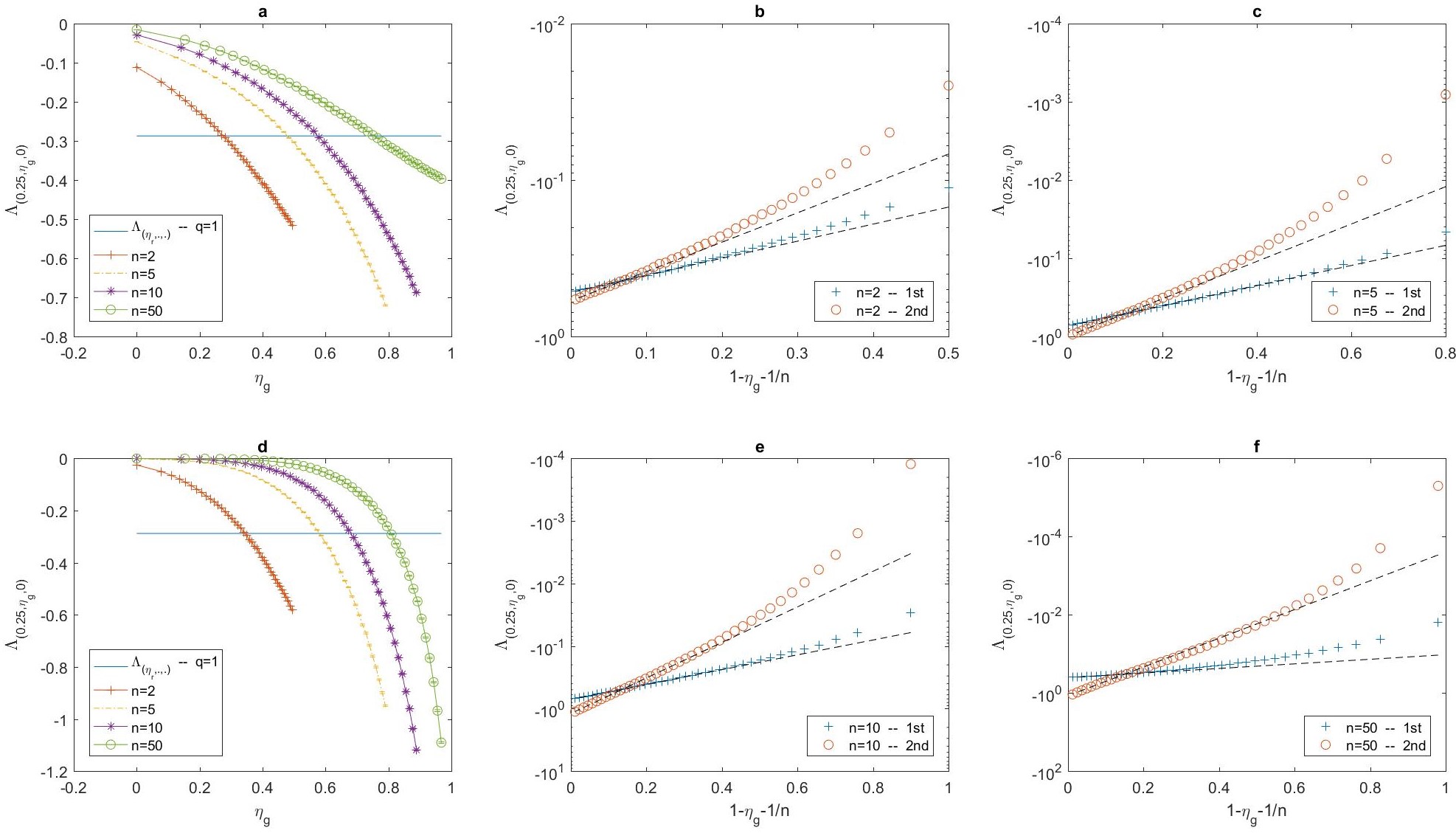}
	\caption{(a), (d): The growth rate on a square lattice with up to $l$th nearest neighbour interactions for a population using a majority rule as its decision making strategy, as a function of $\eta_g$, for different number of states $n$, is plotted. The growth rate of a population with the same representation channel but without communication is plotted as well.  Here $q=1$. In (a) $l=1$ and in (d) $l=2$. We see that the value of communication decreases with increasing noise level and can become negative for high noise levels with a majority rule strategy. Besides the value of communication, increases with increasing $n$. In (a), for $n=50$, we see that the growth rate decreases more slowly with increasing noise level by increasing $n$. The reason is that with higher $n$, the probability that a wrong internal representation is produced due to noise in communication decreases with $n$, and thus communication becomes less misleading in high noise levels when $n$ is larger. (b), (c), (e), and (f): Log linear plot of the growth rate versus $1-\eta_g-\frac{1}{n}$ for different values of $n$. This is the probability by which a correct signal is produced above a uniformly random signal production. We see that the growth rate increases exponentially as the probability of correct signal production increases beyond a uniformly random signal production matrix for all $n$s. The slope of the fit is generally different for different values of $n$.}
	\label{figncPer}
\end{figure*}
\begin{figure}
	\includegraphics[width=1\linewidth]{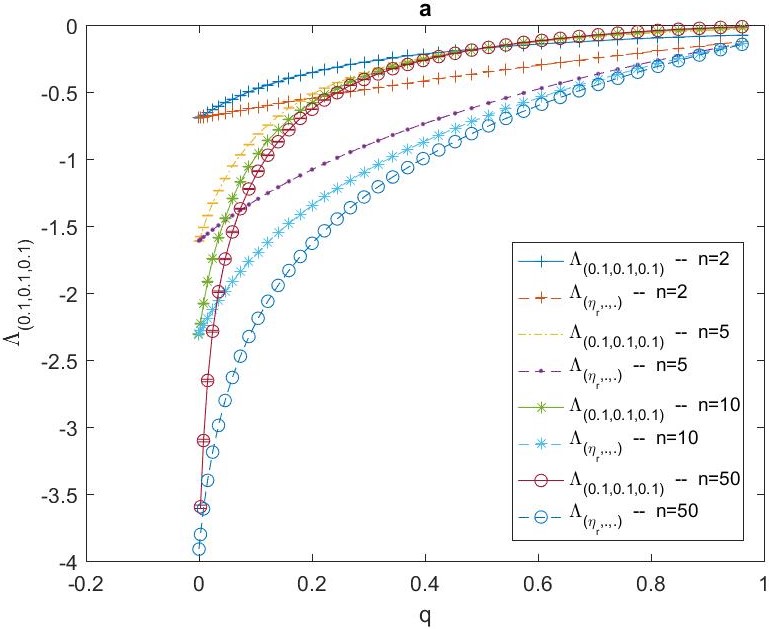}
	\caption{The growth rate on a square lattice for first nearest neighbour interactions versus $q$, for different values of $n$. We see that the value of communication generally increases with $n$.}
	\label{figncq}
\end{figure}
Simulations for local communication done in subsection \secloc, in the main text, and below in this section, are done as follows. We choose the probability of occurrence of each environmental state $p_{\epsilon}$ to be a uniform probability distribution. In each time step, one of the environmental states occurs according to the probability $p_{\epsilon}$. Unless otherwise stated, we set the number of environmental states, internal representations, and signals, to be equal to $n=5$. The population is located on a square lattice. Agents can communicate with their up to $l$th nearest neighbours. $l$ changes in each simulation set as specified in the text. In each environmental step, each agent makes an observation with probability $q$. The observation is made using $R(r|\epsilon)$. This means that as a result of an observation in environment $\epsilon$, an agent obtains a representation $r$. Those who make an observation produce a signal $\sigma$ and transmit it to their up to $l$th neighbours. The receivers transform the received signal $\sigma$ to an internal representation $r^{\rho}$ with conditional probability $C(r^{\rho}|\sigma)$. As a result of observation and communication, each agent reaches an internal state $\pmb{r}$ and chooses its strategy with a majority rule. This means that agents accept the internal representation which happens with highest frequency, and choose its corresponding strategy. If there are ties, that is if there are more than one internal representation with highest frequency, agents randomly choose one of the majority representations. However, if one of the majority representations is supported by personal observation, that is the personal observation of the agent corresponds with one of the representations which happens the highest number of times, the agent chooses the strategy corresponding to that representation. It is easy to see that the growth rate defined in Eq. (\eqlyap) in the main text is equal to:
\begin{align}
\Lambda=\lim_{t \to \infty}\frac{1}{T}\sum_{t=0}^{T-1}\log(d_{t,s_{\epsilon_t}}).
\label{d}
\end{align}
Here, $d_{t,s_{\epsilon_t}}$ is the number of individuals who choose the correct strategy, $d_{t,s_{\epsilon_t}}=\frac{N_{t,s_{\epsilon_t}}}{N_t}$. The quantity which we have calculated, is the same as this, where the time average is done for $T=50$ time steps. The reported growth rates are averages of a sample of $40$ independent runs. and standard errors are calculated as the standard deviation of this sample divided by $\sqrt{40}$.\\
To calculate growth rate according to Eq. (\ref{d}), we need to calculate the number of individuals who choose the correct strategy $d_{t,s_{\epsilon_t}}$. The population size grows according to $N_{t+1}=w_{\epsilon_t}N_{t,s_{\epsilon_t}}$. Hence, in principle we need to update the size of the network according to the current population size in each time step. That is, to calculate $d_{t+1,s_{\epsilon_{t+1}}}$, we need to work with a network of size $N_{t+1}$. However, we note that the communication network affects $d_{t,s_{\epsilon_t}}$, only through its topology and this quantity is independent of the size of the network as long as it is large enough to suppress statistical fluctuations and gives a good estimate of $d_{t,s_{\epsilon_t}}$. This can be seen by referring to Eq. (\ref{het3}) above for population size. We see that it is independent of the size and only dependent on the topology of the network through the degree distribution. Another way to see this is to refer to Eq. (\eqhetsixmain) for long term growth rate in the main text to see that it is independent of the population size and determined by the topology of the communication network. Below, we confirm this independence on size of the network computationally as well. Thus we can fix the network size to a large enough number and estimate $d_{t,s_{\epsilon_t}}$ on this fixed network. In our simulations we have fixed $N=400$ and chosen a $20 \times 20$ square lattice, and calculated $d_{t,s_{\epsilon_t}}$ based on this sample of population of size $N_0=400$.\\ 
In Fig. (\ref{figNq}), we confirm independence of growth rate of the network size. We use Eq. (\ref{d}) to calculate growth rate, by setting $T=50$. The averages and standard errors are calculated based on a sample of size $40$. We calculate growth rate using the algorithm mentioned above, on different network sizes and plot it versus observation probability $q$ in Fig. (\ref{figNq}.a). We see that for low network size $N=25$ extinction occurs. An extinction event in the plot corresponds to at least one extinction event in the sample. However, as long as extinction does not occur, the growth rates calculated on different network sizes are close. As $N$ increases, the result of simulation on different network sizes coincides with a high accuracy. In Fig. (\ref{figNq}.b), we calculate the fraction of individuals who choose the correct strategy, $d_{t,\epsilon_t}$, on different network sizes, as a function of $q$. As seen in Eq. (\ref{d}), this is the only quantity which enters calculation of th long term growth. The reported values and its standard errors are calculated based on a sample of size $T=2000$. As with our parametrization of the CSS matrices, all the states are symmetric, in the sense of having similar conditional distributions, this value does not depend on the environmental state. We see that this number is independent of network size to a high accuracy. This justifies approximating it on a fixed network size.
\subsection{dependence on observation probability}
In simulations in the main text to investigate the dependence of growth rate on noise in representation and production, we fixed the observation probability to $q=1$. In Fig. (\ref{fig23Rq}), Fig. (\ref{fig23Pq}), and Fig. (\ref{figexpq}), we repeat the same experiments for different values of $q$. More precisely, in Fig. (\ref{fig23Rq}), we have fixed $\eta_g=0.1$ and $\eta_c=0$, and plot $\Lambda_{(\eta_r,0.1,0)}$ as a function of $\eta_r$, for different values of $q$. And, in Fig. (\ref{fig23Pq}), we have fixed $\eta_r=0.1$ and $\eta_c=0$, and plot $\Lambda_{(0.1,\eta_g,0)}$ as a function of $\eta_g$, for different values of $q$. We see that all the qualitative conclusions are the same. The value of communication defined as the difference between the growth rate with communication and without communication generally increases as noise in communication decreases, for all values of $q$. In Fig. (\ref{figexpq}), we see that near maximum noise level, the growth rate increases exponentially with noise reduction. However, the slope of the fit depends on $q$: for low $q$s, it increases with increasing $q$, and saturates to a $q$ independent value for high $q$s. Generally, this slope increases with the average number of signals an individual receives through communication. This is so because by increasing the number of signals an individual receive, communication becomes more indispensable and useful and the positive effect of noise reduction manifests itself. This number increases with increasing the local neighbourhood, i.e. increasing $l$, and with increasing $q$. This can be seen in the figures. these are explained in more depth in the main text. Here we see their validity for different values of observation probability.
\subsection{The effect of the number of states}
In Fig. (\ref{figncRep}), Fig. (\ref{figncPer}), and Fig. (\ref{figncq}), we investigate the effect of the number of states $n$. 
As usual we consider a population residing on a square lattice, in which individuals communicate with their up to $l$th nearest neighbours, and use a majority rule as their strategy. We subtract the constant term $\langle \log w_{\epsilon}\rangle$, from the growth rate and plot the remaining entropic part. The set up of the simulations are the same as that described above in the beginning of this section.\\
In Fig. (\ref{figncRep}.a) and Fig. (\ref{figncRep}.d), we plot the growth rate $\Lambda_{(\eta_r,0.25,0)}$, as a function of $\eta_r$, for different number of states $n$. Here $q=1$. The growth rate of a population with the same representation channel but without communication is plotted as well.  In Fig. (\ref{figncRep}.a) $l=1$ and in Fig. (\ref{figncRep}.d) $l=2$. We see that for the chosen noise level in communication , the value of communication is always positive and increases with $l$. For high representation noise level, the growth rate decreases with $n$. This is so because in high representation noise level, the number of individuals who choose the correct strategy approaches $\frac{1}{n}$ which decreases with $n$. In Fig. (\ref{figncRep}.b), Fig. (\ref{figncRep}.c), Fig. (\ref{figncRep}.e), and Fig. (\ref{figncRep}.f), we see the log-linear plot of the growth rate versus $1-\eta_r-\frac{1}{n}$ for different values of $n$s. This is the probability by which a correct representation is formed above a uniformly random representation formation. We see that the growth rate increases exponentially as the probability of correct representation formation increases beyond a uniformly random representation formation matrix for all $n$s. This is the result we saw in the main text. Here, we see it is valid in other values of $n$ as well. The slope of the fit generally increases with $n$.\\
In Fig. (\ref{figncPer}.a) and Fig. (\ref{figncPer}.d), we plot the growth rate, $\Lambda_{(0.25,\eta_g,0)}$ as a function of $\eta_g$, for different number of states $n$.  Here $q=1$. The growth rate of a population with the same representation channel but without communication is plotted as well. In Fig. (\ref{figncPer}.a) $l=1$ and in Fig. (\ref{figncPer}.d) $l=2$. We see that the value of communication decreases with increasing production noise, and can become negative for high noise levels with a majority rule. However, as explained in the main text, the population can correct for this by a simple change of strategy such as ignoring communication and relying on personal observation if made one, in high communication noise regime. We see that for $l=1$ and $n=50$ the growth rate decreases much slower with increasing noise level. This is so because with higher number of states it becomes less probable that any of the wrong representations happens in the majority group in. Consequently, high noise level becomes less detrimental as it becomes less likely that it misleads the individual. However, this probability increases as the number of signals an individual receives, increases. This happens in Fig. (\ref{figncPer}.d), where $l=2$ and individuals receive more signals.  Here we see that in high noise levels growth rate decreases in a steeper way with increasing noise level. In Fig. (\ref{figncPer}.b), Fig. (\ref{figncPer}.c), Fig. (\ref{figncPer}.e), and Fig. (\ref{figncPer}.f), we plot the growth rate versus $1-\eta_g-\frac{1}{n}$ in a log-linear plot, for different values of $n$s. This is the probability by which a correct signal is produced above a uniformly random signal production. We see that the growth rate increases exponentially as the probability of correct signal production increases beyond a uniformly random signal production matrix, for all $n$s. The slope of the fit is generally different for different values of $n$. However, this slope is smaller for higher number of states. As mentioned, the reason is that in high noise level, the probability that a wrong representation is produced due to noise in communication becomes smaller with higher number of states, and consequently communication becomes less misleading and the growth rate decreases more slowly with increasing noise level.\\
In Fig. (\ref{figncq}) We plot the growth rate $\Lambda_{(0.1,0.1,0.1)}$ as a function of $q$ for different $n$s, setting $l=1$. We see in this low level of noise in communication, the growth rate is always above that of the same population without communication and increases with $q$ more rapidly with higher $n$ values.

\end{document}